\documentclass[pre,showpacs,preprintnumbers,amsmath,amssymb]{revtex4-1}
\usepackage{graphicx}
\usepackage{bm}
\usepackage{color}
\usepackage{amssymb}
\usepackage{epsfig}
\DeclareGraphicsExtensions{.png,.pdf}

\begin{document}

\title{Average Spectral Density of  Multiparametric Gaussian Ensembles of Complex Matrices}
\author{Mohd. Gayas Ansari and Pragya Shukla$^*$}
\affiliation{ Department of Physics, Indian Institute of Technology, Kharagpur-721302, West Bengal, India \\\\
{\it Corresponding Author E-Mail: shukla@phy.iitkgp.ac.in}}
\date{\today}

\widetext

\begin{abstract}

A statistical description of part of a many body system  often requires a non-Hermitian random matrix ensemble with nature and strength of randomness sensitive to underlying system conditions.  For the ensemble to be a good description of the system, the ensemble parameters must be determined from the system parameters. This in turn makes its necessary to analyze a wide range of  multi-parametric ensembles with different kinds of matrix elements distributions. The spectral statistics of such ensembles is not only  system-dependent but also non-ergodic as well as non-''stationary''.   

A change in system conditions can cause a change in the ensemble parameters resulting an evolution of the ensemble density and it is not sufficient to know the statistics for a given set of system conditions. This motivates us to theoretically analyze  a multiparametric evolution of the ensemble averaged spectral density of a multiparametric Gaussian ensemble on the complex plane.  Our analysis reveals the existence of an evolutionary route common to the ensembles belonging to same global constraint class and thereby derives a complexity parameter dependent formulation of the spectral density for  the non-equilibrium regime of the spectral  statistics,  away from Ginibre equilibrium limit.

\end{abstract}

\maketitle

\section{Introduction}

The unavoidable contact with environmental or dissipative aspect of real physical systems makes it necessary to consider the physics of non-Hermitian generators of the dynamics. Indeed recent developments in a  wide range  of complex systems e.g. dissipative quantum systems \cite{berry, r1, r2, r3, r4, r5, r6, r7, r8, r9, r10, r11, r12, r13, r14, r15, r16, r17, r20, r21, r22, r23}, neural network dynamics and biological growth problems \cite{sc, ns, kr1, ahn, afm},statistical mechanics of flux lines in superconductors with columnar disorder \cite{fte} classical diffusion in random media \cite{cw} have refocused the attention on random non-Hermitian operators and a search for the mathematical models for their statistical properties e. g. \cite{cw, cb,  psnh, ben1, ben0, ben2, most, jk, gnt, ir1, ir2, fh1, fz, fy1, fy2,gin, fy3, nh, prosen, hkku,tk,bp, gnv, swhjy, mcn,  nm}.  As the average behavior of many of the physical properties can be determined from  the average spectral density of the operators,  its theoretical formulation is very desirable and is the primary focus of present study.

In general, a real physical system under consideration is often complex.   Enormous success of Hermitian random matrix ensembles  in past with the Hermitian operators of complex systems  \cite{haak, meta,fkpt5,  brody,  apbe, psrev, gmw, psall, psall1, psco, ptche, apce, psalt, psacu, psflat} encourages us to seek non-Hermitian random matrix ensembles as the models for the non Hermitian operators.  
 But, as in  the Hermitian case, the manifestation of the complexity is in general system dependent and the choice of an appropriate ensemble representing  the system requires a prior knowledge of the relations among ensemble and system parameters and  imposes various matrix as well ensemble constraints  \cite{psrev}. This in turn requires an analysis of a wide range of system dependent random matrix ensembles, with matrix elements described by multi-parametric distributions  which may also be correlated.   The growing applications of such ensembles in complex systems has in past led to  development of a number of theoretical tools  e.g. orthogonal polynomial method \cite{fy1, fy2, fy3, gnv},  Green's function approach \cite{fh1, nh,  sc},  cavity method \cite{mcn, nm},  nonlinear sigma model approach \cite{haak} etc.  The results obtained are however often based on system specific approximations   used to simplify the analysis \cite{gin, haak, bp, gnv, tk, mnr}.  
This motivated one of us in \cite{psnh} to seek a theoretical formulation where all system information manifest  through a single function of all system parameters.  The study \cite{psnh} however did not pursue a detailed analysis of the formulation e.g analysis of the spectral density or its fluctuations.  Our primary objective in the present work is to bridge this information gap by deriving a common mathematical formulation for the spectral density of a wide range of Gaussian ensembles of  complex matrices with varying degree of sparsity.

The contact with the environment subjects a physical system to many perturbations often causing a variation of system parameters with time.   With a non-Hermitian Hamiltonian basically describing the  interaction between system and its environment, and the latter often being dynamic, the Hamiltonian matrix is expected to evolve in complex matrix space with time-varying system conditions.   But as the Hamiltonian matrix can also be represented by an ensemble or complex matrices at each instant of time; a variation of ensemble parameters results in evolution of the ensemble density on the ensemble space.  This give rise to a natural query whether it is possible to find a  path in the ensemble space, the evolution along which  can exactly be mapped to evolution on the complex matrix space.   The existence of such a path implies that the system is appropriately represented by the same ensemble for all instants of time.   Additionally while the evolution in matrix space is  governed by many system parameters,  all of them vary with time, thereby implying time as the ultimate independent variable.  The evolution in ensemble space is however governed by  many ensemble parameters.  It is therefore desirable to seek  a  time-like single parametric evolution in ensemble space too.  Such an evolution if possible would  indicate existence of universality classes, different from that of Ginibre and  among  non Hermitian ensembles belonging to a wide range of parametric dependences.
Relevant initial steps   in this context  were made in the study \cite{psnh}; the latter derived, by an exact theoretical route, a diffusion equation for the joint probability distribution function ({\it jpdf}) for the eigenvalues of a multi-parametric Gaussian ensemble of non-Hermitian matrices. While a random perturbation of the matrix elements of such an ensemble can subject them to diffusive dynamics at different parametric scales (given by ensemble parameters), the diffusion of the eigenvalues was shown to be governed by  a single  functional of all the ensemble parameters. As the latter are functions of the system parameters, the functional, referred as the complexity parameter, contains the relevant information about the system under consideration. The existence of such a formulation, verified by the detailed numerical analysis on many body as well as disordered systems, has already been shown in past for the Hermitian operators \cite{psco, psalt, psall, psall1, ptche, psacu}.  As mentioned above,  while the basic formulation for the  non-Hermitian cases was derived in \cite{psnh}, its detailed analysis on the complex plane e.g. the statistical fluctuations of the radial and angular parts of the  eigenvalues, as well as the desired numerical verification was not pursued earlier (mainly due to lack of interest in non-Hermitian operators in past) and has been missing so far.  An intense interest in the domain recently as well as  robustness and wider applicability  of the formulation now  makes it desirable to extend our analysis of \cite{psnh} further.  In  the present study,  we derive an explicit formulation of the ensemble averaged spectral density of the sparse complex random matrices as a function of the complexity parameter. Our approach is based on four crucial steps: (i) first considering the evolution of a  multiparametric Gaussian ensemble of complex matrices in the ensemble space due to variation of ensemble parameters and described by a combination of first order parametric derivatives, (ii) showing that it can exactly be mimicked by a diffusion in the matrix space with a constant drift, (iii) the multiparametric evolution can be reduced to an evolution governed by a single function of all  ensemble parameters (therefore referred as complexity parameter),   (iv) solving evolution equation  for the average density of states.

The paper is organized as follows. Section II  introduces the complexity parametric formulation  of the ensemble density and thereby that of the {\it jpdf} of the eigenvalues on the complex plane. While the latter was derived in \cite{psnh} by a direct route,  the derivation presented here is based on the ensemble density and  is  richer as it can be extended to derive the distribution of the eigenfunctions too.  
 As previous studies of the non-Hermitian ensembles  are often focussed on those with free of any parameters e.g. Ginibre ensemble \cite{haak}, the available results correspond to ergodic limit of statistical fluctuations. The presence of many parameters usually makes  the spectral statistics non-ergodic as well as non-''stationary''.  As our analysis is focussed on multiparametric ensembles, it is necessary to first define the concept of non-ergodicity as well as non-''stationarity'' in an ensemble; this is discussed in section III.  Section IV  presents a derivation of the diffusion equation for  the ensemble averaged spectral density in polar coordinates and its solution.  
This is followed by the numerical analysis, in section VI,  for three different multi-parametric Gaussian ensembles of non-Hermitian matrices.  The choice of these ensembles is basically motivated from their applicability to real systems as well as from the already available results for the similar Hermitian ensembles that helps in the comparative studies of the spectral fluctuations under Hermitian and non-Hermitian constraint.  We conclude in section VII with a brief discussion of our main results, insights and open issues for future.    We also note that  a given symbol with different number of  subscripts has different meaning throughout the text (e.g.  $a_{\mu \nu}$ and $a_{\mu}$ are not same).

\section{Diffusion of the eigenvalues on the complex plane}

Consider a complex system represented by a non-Hermitian matrix with known constraints only on the mean and  variances of the matrix elements and their pair wise correlations. 
Based on the maximum entropy hypothesis (MEH), it can be described by an ensemble of $N\times N$ complex matrices $H$ defined by a multiparametric Gaussian ensemble density $\tilde\rho (H,y,x)  \propto  {\rm exp}\left[- H_v^{\dagger}  {\mathcal V} H_v\right]$  with $H_v$ as the column vector consisting of all matrix elements $H_{kl;s}$ and ${\mathcal V}$ as the covariance matrix.  
However,  for technical reasons described in next section,  here we consider an equivalent but alternate form,  namely, $\tilde\rho (H,y,x) = C \rho(H,y,x)$ with
\begin{eqnarray}
 \rho(H,y,x) = {\rm exp}\left[
- \sum_{s=1}^\beta \sum_{k, l}
 (y_{kl;s} H_{kl;s}^2 + x_{kl;s} H_{kl;s} H_{lk;s})\right] 
\label{pdf}
\end{eqnarray}
 with $\beta=2$, $C$ as a constant determined by the normalization condition $\int  \tilde\rho \, {\rm d}H =1$,    $y$ and $x$ as the sets of distribution parameters $y_{kl;s}, x_{kl;s}$ of  various matrix elements.  Here the subscript $s$ on a variable refers to  one of its components i.e real ($s=1$) or imaginary ($s=2$) part, with $\beta$ as
 total number of the components.  
 As the ensemble parameters $y_{kl;s},  x_{kl;s}$ can be arbitrarily chosen (including an infinite value for non-random entries), $\rho(H,y,x)$  can represent  a large class of non-Hermitian matrix ensembles e.g. varying degree of sparsity or bandedness.   For example,  even a simple choice of almost all $y_{kl;s} \rightarrow N/(1-\tau^2)$ and $x_{kl;s} \rightarrow (-1)^{s-1} N \tau/(1-\tau^2)$ with $\tau \rightarrow 0, \; \pm 1$ can  lead to three different universality classes of stationary ensembles
namely,  Ginibre ($\tau=0$),  GUE ($\tau=1$) and  the ensemble of complex antisymmetric matrices or GASE ($\tau=-1$) and many non-stationary ones intermediate between them for $0 < |\tau | < 1$  \cite{sc,  fy1, nh}.    Further with choice of  $y_{kl;s}, x_{kl;s}$ as the functions of system parameters, $\rho$ can act as an appropriate  model  for a wide range of non-Hermitian complex  systems \cite{psnh}.

\subsection{Evolution of the ensemble density}

A change in system conditions can affect the matrix elements $H_{kl}$,  resulting in evolution of $\rho$ in matrix space.  But as ensemble parameters are functions of system parameters,  they also vary, leading to an evolution of $\rho$ in ensemble space too.  Here the matrix space is a $2 N^2$ dimensional  space  (consisting of  $N^2$ real and $N^2$ imaginary components of the matrix elements,  with each point in the space corresponds to a $N \times N$ non-Hermitian matrix. The ensemble space   is a $2 N^2$ dimensional  space consisting of independent ensemble parameters $x_{kl}, y_{kl}$ with each point representing the state of the ensemble.

For the evolving ensemble to continue representing the system,  it is desirable that the evolution of $\rho$ in matrix space  is exactly mimicked by that in ensemble space.  This motivates us to consider a first order variation  of the ensemble parameters $y_{kl;s} \rightarrow y_{kl;s}+\delta y_{kl;s}$ and $x_{kl;s} \rightarrow x_{kl}+\delta x_{kl;s}$  over time and seek (i) what type of matrix space dynamics they correspond to, and, (ii) can it lead to an equilibrium ensemble with stationary correlations? More explicitly, we consider following combination of the first order parametric derivatives

\begin{eqnarray}
T \; \rho \equiv \sum_{k,l}^N \sum_{s=1}^{\beta}  \left[ A_{kl;s} {\partial  \rho \over\partial y_{kl;s}}+ B_{kl;s}  {\partial  \rho\over\partial x_{kl;s}}\right] 
 \label{trho}
\end{eqnarray}  
with
\begin{eqnarray}
A_{kl;s} &=& 2 y_{kl;s} \left(\gamma - 2 y_{kl;s} \right) -x^2_{kl;s}  
 \nonumber  \\
B_{kl;s} &=& 2 x_{kl;s} \, \left(\gamma - 4 \, y_{kl;s}\right)
\label{dab}
\end{eqnarray} 
%
%

Here $\gamma$ is an arbitrary parameter that gives the freedom to choose the end point of the evolution.   
For brevity of notations,   hereafter $\sum_{k,l}^N \sum_{s=1}^{\beta}$ will be written just as $\sum_{k,l;s }$.

With help of its Gaussian form,  the multi-parametric evolution of $\rho$ can  be described exactly  in terms of a diffusion, with finite drift,  in $H$-matrix space,

\begin{eqnarray}
T \rho + C_1 \, \rho
=\sum_{k,l;s}
{\partial \over \partial H_{kl;s}}\left[
 {\partial \over \partial  H_{kl;s}}  +\gamma \; H_{kl;s}  \right] \rho
\label{td1}
\end{eqnarray}
with $C_1= \sum_{k,l;s} \left(\gamma - 2 \; y_{kl;s} \right)$.

 As eq.(\ref{td1})  is difficult to solve for generic parametric values,  we seek a transformation from the set of $M$ parameters $\{y_{kl;s}, x_{kl;s} \}$ to another set $\{t_1,\ldots, t_M \}$,  with  $M = 2 N^2$, 
such that only $t_1$ varies under the evolution governed by the operator $T$ and rest of them i.e $t_2, \ldots, t_M$ remain constant: $T \rho+ C_1 \, \rho\equiv \frac{\partial \rho}{\partial t_1}$ or $T \rho_1  \equiv {\partial \rho_1\over\partial t_1}$ 
where  $\rho_1 = C_2 \; \rho$   with $C_2$ defined by the condition $T C_2 -C_1 C_2 ={\partial C_2 \over \partial Y}= 0$ (discussed in  {\it appendix A}).  
This implies

\begin{eqnarray}
 {\partial \rho_1\over\partial t_1} &=&
\sum_{k,l;s}
{\partial \over \partial H_{kl;s}}\left[
 {\partial \over \partial H_{kl;s}} +\gamma \; H_{kl;s}  \right] \rho_1 
 \label{rhot1}
\end{eqnarray}
 Clearly $t_k$  should satisfy the condition that
 
\begin{eqnarray}
   \sum_{k,l; s}
 \left[ A_{kl;s} {\partial t_k \over\partial y_{lk;s}}
 +  B_{kl;s}  {\partial t_k \over\partial x_{kl;s}}\right]
 = \delta_{k1}  
\label{ttk}
\end{eqnarray}

As discussed in {\it appendix A}, the above set of $M$ equations can be solved by standard method of characteristics.  The solution for $k=1$  is 
\begin{eqnarray}
t_1 &=&  \pm {2\over  M} \sum_{k,l;s}
\int \frac{{\rm d}y_{kl;s}}{y_{kl;s} \, w} + c_0  
\label{y}
\end{eqnarray}
with  $c_0$ as the constant of integration, $w={2 \gamma - 4 y_{kl;s}-x_{kl;s}^2 }$ and $x_{kl;s}$ expressed as function of $y_{kl;s}$. 
Here the sign for $t_1$ is chosen  so as to keep it  increasing above the  initial value (following standard diffusion equation approach where $t_1$ plays the role of time and is always positive).    

We note that the form of $t_1$ in eq.(\ref{y}) is obtained by assuming  all $y_{kl;s} \not=0$.  For $y_{kl;s}=0$,   eq.(\ref{dab}) gives $A_{kl;s}=0$ and the corresponding derivative do not contribute to eq.(\ref{trho}).   Indeed,  for the  cases in which a specific parameter $y_{kl;s}=0$ throughout the evolution (with $k, l, s$ arbitrary),  the  number $M$ of evolving parameters  is  less than $2 N^2$.

 For the case $x_{kl;s} \not= 0$, the constants  $c_{kl;s}$ and $\tilde c_{kl;s}$ are given by the relations  $x_{kl;s} = \tilde c_{kl;s} (y_{kl;s} -y_{lk;s}) (\gamma -2 y_{kl;s} - 2 y_{lk;s})$ and $y_{lk;s} = c_{kl;s} \; y_{kl;s}$.  Integration in eq.(\ref{y}) further gives

\begin{eqnarray}
t_1 =  \pm {2\over  \gamma M} \sum_{k,l;s}
\int \frac{{\rm d}y_{kl;s}}{y_{kl;s} \, (2 \gamma - a_1 y_{kl;s} + a_2 y_{kl;s}^2 -a_3 y_{kl;s}^3)}  + c_0.
\label{yt1} 
\end{eqnarray}
with $a_1=4+\gamma^2 \tilde c_{kl;s}^2 (1- c_{kl;s})^2$, $a_2=4 \gamma \tilde c_{kl;s}^2 (1+c_{kl;s}) (1- c_{kl;s})^2 $, $a_3=4 \tilde c_{kl;s}^2  (1- c_{kl;s}^2)^2$. 
Here the symbol  $\sum_{k,l;s}'$ indicates the summation over non-zero $y_{kl;s}$ only.

Similarly solution of eq.(\ref{ttk}) for $k >1$  leads to $M-1$ constants of integrations which can be determined by the initial conditions as well as global constraints on the dynamics. (As a similar approach for Hermitian operators has been discussed in a series of studies \cite{psall}, and in \cite{psnh} for non-Hermitian operators, the details of the derivation are not included in this work). To distinguish it from $t_k$ with $k >1$,  hereafter $t_1$ will be referred as $Y$.

As a simple example (also used  in our numerics discussed in section V), here we consider the case with $x_{kl;s}=0$.  
 The latter corresponds to $\tilde c_{kl;s}=0$ and thereby $w=2(\gamma -2 y_{kl;s})$. Following from eq.(\ref{y}), we then have
\begin{eqnarray}
Y \equiv t_1 =  \pm {2 \over M}  \sum_{k,l,s}'  {\rm ln} \; {y_{kl;s}\over |\gamma-2 y_{kl;s}|}  + c_0,
\label{yt1b}
\end{eqnarray}
and $t_k =c_k$ for $k >1$ with $c_k$ as arbitrary constants of integration.  
(This can also be seen directly from eq.(\ref{ttk}): for $x_{kl;s}=0$,   it reduces to the condition $ \sum_{k,l, s} ' y_{kl;s} \; {\partial t_k \over\partial y_{kl;s}} = \delta_{k1}$ which on solving gives eq.(\ref{yt1b}) for $Y$).

An important insight in eq.(\ref{rhot1})  can be gained by noting the similarity of its form with the standard Fokker Planck (FP) equation,  describing the diffusive dynamics  of  $2 N^2$ "particles" $H_{kl;s}$   in "time" $t_1  \equiv Y$  in a harmonic external potential and subjected to white noise.   As the  standard FP formulation is based on consideration of positive time like variable,  hereafter we will consider  $ Y \ge 0$; the latter is  ensured by choosing the appropriate sign in  eq.(\ref{yt1}) or eq.(\ref{yt1b}).  
Similar to FP equation,  $\rho_1$ approaches  the steady state in the limit ${\partial \rho \over \partial Y} \to 0$  of eq.(\ref{rhot1})  or as $ Y \to \infty$.   For example,  
a choice of almost all $y_{kl;s} \rightarrow N/(1-\tau^2)$ and $x_{kl;s} \rightarrow (-1)^{s-1}  \tau N/(1-\tau^2)$  with $\gamma=1$ fulfills the condition for $\tau \rightarrow 0, \; \pm 1$  and   can therefore lead to three different types of steady states, namely,  Ginibre ($\tau=0$), GUE ($\tau=1$) and   GASE ($\tau=-1$).   As expected,  the solution of eq.(\ref{rhot1}) with its left side set to zero is consistent with already known distributions for Ginibre,  GUE and GASE (also follows from eq.(\ref{pdf}) on substitution of the above values of $x_{kl;s}, y_{kl;s}$).
 As mentioned previously,  for a random matrix  ensemble to represent a complex system,    an appropriate choice of the ensemble parameters must take  into account all system-specifics,   thereby rendering them to be  functions of the system parameters. This in turn makes $Y$ as the functional of the latter and  thereby  a measure of the complexity of the system.

 As mentioned below eq.(\ref{dab}), $\gamma$ is an arbitrary parameter and can be set to unit value too.  Nonetheless it is useful to retain $\gamma$ in the formualtion for two reasons: (i) it gives the freedom to analyze the statistical behavior of the system   in $\gamma=0$ or $\infty$ limit (e.g.  the dynamics in absence  off the confining  harmonic potential part or in reducing the role of diffusion relative to drift term), (ii) $\gamma$ also plays the role of a spectral unit in numerical analysis (e.g. while analyzing the approach  of different ensembles to  Ginibre universality class).

We emphasize that the form of eq.(\ref{rhot1})  is based on  the  specific forms for $A_{kl;s}$ and $B_{kl;s}$ given in eq.(\ref{dab}).  Indeed it is motivated by our search for a combination of the parametric derivatives that would lead to  a Dyson  Brownian dynamics model in non-Hermitian matrix space  (introduced by Dyson for Hermitian case and discussed in detail e.g. in \cite{haak, meta}),  starting from an arbitrary initial condition and with a Ginibre ensemble as the equilibrium limit. The search  is motivated  in turn by a hope to connect and utilise already existing  technical tools developed for Dyson Brownian ensembles of Hermitian matrices \cite{haak, meta, apbe, psrev,  circ, psall, psall1, psco}.    

The above mentioned connection  motivates a natural  query as to what type of dynamics  $\rho$ (eq.(\ref{pdf})) undergoes if  subjected to an evolution governed by  an arbitrary combinations of parametric derivatives? Alternatively,  if we consider a Brownian dynamics model for a non-Gaussian ensemble (i.e the combination of matrix element derivatives in right side of eq.(\ref{rhot1})),  can it still be described by a combination of the first order or higher order parametric derivatives?  Some insights in this context are discussed in previous studies on Hermitian matrix ensembles \cite{psalt, psco}.
As discussed in section V of \cite{psalt}, not all parametric combinations would lead to an evolution in the Hermitian matrix space with an equilibrium limit.  Besides,  as discussed in section II.C or II. D of \cite{psco},  not all combinations seem to be easily reducible to a single parametric dependence of the evolution (although a detailed investigation is needed in this context).

\subsection{Evolution of the {\it jpdf} of the eigenvalues}

Contrary to a spectral degeneracy of a Hermitian matrix,  a nonhermitian matrix is non-diagonalisable at the exceptional points i.e  the points in the parameter space at which some of the eigenvalues become degenerate and the corresponding eigenvectors coalesce into one \cite{ir2}.   But as the number of  exceptional points is at most finite,  the non-Hermitian matrices are considered to be diagonalisable.  We hereafter refer the parametric conditions avoiding exceptional points as non-exceptional.

Assuming non-exceptional parametric conditions,  a non-Hermitian matrix can be diagonalized by a transformation of
the type $Z = U H V$ with $Z$ as the matrix of eigenvalues
 $z_j $ and $U$ and $V$ as the left and right eigenvector matrices
respectively.  Here we consider the case of complex matrices  ($\beta=2$) only, with its eigenvalues
$z_j \equiv \sum_{s=1}^2 (i)^{s-1} z_{js} $, in general,  distributed over an area in the complex plane and $U, V$ as unitary matrices; 
here subscript  $s$ refers to the real or imaginary components 
of the eigenvalues). Let  $\tilde P(z,y,x)$ 
be the {\it jpdf} of the eigenvalues  at arbitrary  values of  $y_{kl;s}$ and $x_{kl;s}$ with $z \equiv \{z_i\}$. 
An integration of eq.(\ref{rhot1}) over all eigenfunctions components leads to  an evolution of $\tilde P(z,y,x)$ with $t_1$;  for symbols consistency with earlier works,  hereafter $t_1$ will be referred as $Y$.  
Let $P_1(z_1, \ldots, z_N)$ be related to the normalized distribution by $\tilde P = C_2 P_1/C$, its $Y$ governed evolution can then be  described  as (details discussed in {\it appendix B}) 
 \begin{eqnarray}
{\partial P_1\over\partial Y} &=&\sum_{s=1}^2 \sum_{n=1}^N{\partial \over \partial z_{ns}}\left[{\partial \over \partial z_{ns}}  - 2 \, {\partial {\rm ln} |\Delta_N (z)| \over \partial z_{ns}}+ \gamma z_{ns} \right] P_1
\label{pcmp}
\end{eqnarray}
with $\beta=2$ and 
\begin{eqnarray}
\Delta_N (z) \equiv \prod_{k < l}^N  (z_k-z_l)
\label{delta}
\end{eqnarray}

As discussed in {\it appendix B}, the derivation of eq.(\ref{pcmp}) is based on the set of eq.(\ref{evf}) describing the response of the eigenvalues as well as the eigenvectors to a small change in the matrix element $H_{kl}$; the assumption of $U, V$ as unitary matrices is necessary for these relations. As discussed in \cite{psnh}, an assumption of $U V=I$ only leads to different dynamics of eigenvalues and eigenfunctions with varying matrix elements; this in turn leads to a n evolution equation for $P_1$ different from eq.(\ref{pcmp}). 




In the limit  ${\partial P_1\over\partial Y} \rightarrow 0$ or 
$Y\rightarrow \infty$, the  evolution described by eq.(\ref{pcmp}) reaches the equilibrium steady state of the Ginibre ensemble with  $ P_1(z; \infty) \, \equiv |Q_N|^2 = \prod_{j< k}|\Delta_N (z)|^2{\rm e}^{-{\gamma\over 2}\sum_k |z_k|^2}$; the steady state limit occurs for the system conditions resulting in  almost all $y_{kl;s} \rightarrow N$ and $x_{kl;s} \rightarrow 0$ with $\gamma=1$. Thus eq.(\ref{pcmp}) describes a cross-over from a given initial ensemble
(with $Y=Y_0$) to Ginibre ensemble as the equilibrium limit with $Y-Y_0$ as the
crossover parameter. The non-equilibrium states of the crossover, given by non-zero finite values of $Y-Y_0$, are various ensembles of the complex matrices corresponding to varying values of
$y_{kl}$'s and $x_{kl}$'s thus modelling different complex systems.  A point worth emphasizing here is following: while the "equilibrium" and "non-equilibrium" states  mentioned above occur on the ensemble space,  such ensembles may well-represent complex systems in e.g. thermal non-equilibrium stages too.

\section{Local ergodicity and ''non-stationarity'' of the spectrum}

The spectral density $\rho(z)$ at a  point $z$ on the complex spectral plane is defined as 

\begin{eqnarray}
\rho(z)  = \sum_{k=1}^N \delta^2(z-z_k)
\label{rho1}
\end{eqnarray}
 where $\delta^2(z-z_k) \equiv \delta(z-z_k) \delta(z^*-z_k^*)$ with $z^*$ as complex conjugate of $z$ subjected to the normalization condition $\int \rho(z)  \, {\rm d}z \, {\rm d}z^* =1$.  Its average over an  ensemble  at a given point $z$ on the spectral plane can then be expressed as    

 \begin{eqnarray}
\langle \rho(z) \rangle =    \int \delta^2 (z-z_1)  P(z_1,\ldots, z_N) \, {\rm  D}\Omega
\label{rn}
\end{eqnarray}
with ${\rm D} \Omega \equiv  \prod_{k=1}^N \;{\rm d}z_{k} \; {\rm d}^*z_k $ and $\langle . \rangle$ as the ensemble average,

Under assumptions that $\rho(z)$  is separable into two components, a slowly varying $\rho_{sm}$ (the secular behavior), and a rapidly varying $\rho_{fluc}$, it can be written as $\rho =\rho_{sm}(z) + \rho_{fluc}(z)$ (describing the local  random excitations of $\rho_{sm}$).  A spectral averaging over a scale  larger  than that of the  fluctuations gives $ \int_{z-\Delta z}^{z+\Delta z}  \; {\rm d} z \; {\rm d}z^* \; \rho_{fluc}(z) =0$ and  $\rho_{sm}(z) = \frac{1}{4 \Delta z \Delta z^*} \int_{z-\Delta z}^{z+\Delta z} {\rm d} z \,\int_{z^*-\Delta z^*}^{z+\Delta z^*} {\rm d} z^*\, \rho(z)$. The removal of the slowly varying part $\rho_{sm}$ by a rescaling then provides the information about  local fluctuations.  Referred  as  "unfolding" of the spectrum, this is achieved by redefining the real and imaginary parts of the spectral variable $z$: 
$e_r=\int^{z_r} \sqrt{{\rho_{sm}}(z) }\; {\rm d}z_r$ and $e_i=\int^{z_i} \sqrt{{\rho_{sm}}(z) }\; {\rm d}z_i$ of the eigenvalues $e_n$ which results in a uniform local mean spectral density \cite{psnh, psgs2}.

\subsection{Local ergodicity of the spectrum}

Experimental results are often based on a single spectrum analysis; this leaves the unfolding by $\rho_{sm}(z)$  as the only option. But theoretical   modelling of a single  complex system by an ensemble of its replicas is often based on the assumptions of local ergodicity i.e the averages of its properties over a given spectral range, say $\Delta^2 z \equiv \Delta z \Delta z^*$ around $z$, for a single matrix  (referred as the spectral average) are same as the averages over the ensemble  at a  fixed $z$ (referred as the ensemble average).  Mathematically this implies 

\begin{eqnarray}
\langle \rho_{sm}(z) ^2 \rangle - \langle \rho_{sm}(z) \rangle^2 \rightarrow 0,  \qquad \qquad \langle \rho_{sm}(z) \rangle= \langle \rho(z) \rangle
\label{erg}
\end{eqnarray}
with $\langle . \rangle$ implying an ensemble average for a fixed $z$
and thereby permits it to be used for a theoretical unfolding of the spectrum. Prior to application of theoretical results to experimental data,  it is therefore necessary to check the ergodicity of the spectrum. (An important point worth emphasizing here is as follows. The concept of ergodicity, a standard assumption in  conventional statistical mechanics,  implies  analogy of the phase space averaging (i.e averaging over an ensemble of micro states) of a measure with time averaging. In case of a random matrix ensemble, the phase space averages  are replaced by the averages over an ensemble of matrices and time averages by the spectral averages; the ergodicity of a measure now implies the analogy of ensemble averaging with  spectral averaging. )

The fluctuations in local spectral density can in principle be measured in terms of the  $n^{\rm th}$ order spectral density correlation $R_n (z_1,\ldots, z_n; Y)$, defined as \cite{psnh}

 \begin{eqnarray}
 R_n = { N! \over {(N-n)!}}\int P(z_1,\ldots, z_N; Y) \, {\rm  D}\Omega_{n+1}
\label{rn0}
\end{eqnarray}
with ${\rm D} \Omega_{n+1}\equiv  \prod_{k=n+1}^N \; {\rm d} z_k \, {\rm d}z^*_{k}$. 
The above definition is valid in general for an arbitrary non-Hermitian ensemble.
As clear from a comparison with  eq.(\ref{rn}), the case $n=1$ corresponds to  the ensemble averaged spectral density $R_1 = N \, \langle \rho \rangle$  and is subjected to  following normalization condition:  $\int {\rm d}z \; {\rm d}z^* \, R_1(z) = N$.

An ergodic level density  does not by itself implies the ergodicity of the local density fluctuations.  Here we generalise the definition for the non-Hermitian case:  the ergodicity   of a local property say "$f$" can be defined as follows: let $P(z, s)$ be the distribution function of $f$ in the neighbourhood ''s'' of an arbitrary  region "$z$" over the ensemble and ${\tilde P}_k(z, s)$ be the distribution function of $f$ over a single realisation, say "$k^{th}$" of the ensemble: $f$ is defined as ergodic if it satisfies the following conditions:
(i) the distribution for almost each realization is same i.e  ${\tilde P}_k(z, s)$ is independent of $k$ for almost all $k$, implying ${\tilde P}_k(z, s) = {\tilde P}(z, s)$, 
(ii) the distribution over the ensemble  is analogous to that of single realization $P(z, s) = {\tilde P}(z, s)$, 
(iii) $P(z,s)$ is stationary on the spectral plane i.e independent of $z$: $P(z, s) = P(z', s)$  for arbitrary points $z, z'$ on the complex plane,
However if condition (iii) is not fulfilled,  $f$ is then referred as locally ergodic. More clearly, an ensemble is locally ergodic in a spectral range say $\Delta z$ around $z$, if the averages of its local properties over the range for a single matrix  (referred as the spectral average) are same as the averages over the ensemble  at a  fixed $z$ (referred as the ensemble average).  It is worth re-emphasizing here the relevance of the translational/ rotational invariance in the bulk of a spectrum: as clear from the above, it implies  an invariance of the local fluctuation measures from one spectral point on the complex plane to another. A non-invariant spectrum can at best be locally ergodic and therefore a complete information about the spectrum requires its analysis at various spectral ranges.

\subsection{Non-stationarity of the spectrum}

 A generic random matrix ensemble need not contain an explicit time dependence and it is necessary to first  define ''stationarity'' in this context. This refers to a  dynamics of  the ensemble, in ensemble space, due to variation of the ensemble parameters; this in turn leads to leads to variation of the statistical properties. 
The stationarity (equilibrium) limit here corresponds to the ensemble parameter values for which ensemble becomes invariant under a set of  transformations related to global constraints. The intermediate states of the ensemble as its parameters vary from an arbitrary initial state to stationary limit, are then referred as ''non-stationary'' or ``non-equilibrium'' ensembles.

As clear from the above,  a non-stationarity of the spectrum in present context refers to its pseudo time-dependence. For example, as mentioned above, the solution of eq.(\ref{pcmp})  corresponds to stationary Ginibre ensemble in the equilibrium limit $Y \to \infty$. The intermediate states of the evolution for finite $Y$ are then referred as non-stationary ensembles. It is important to reemphasize the difference between non-ergodicity and non stationarity of the spectrum: the former refers to a non homogeneous statistics on the complex spectral plane at a fixed $Y$, the latter refers to change of statistics at a specific spectral point of interest  with changing $Y$. 

 As discussed in previous section,   the ensemble density  (\ref{pdf}) is both non-ergodic as well as non-stationary.  In the next section, we analyze the dynamics of its spectral density in complex plane as $Y$ varies.

\section{Spectral Density on the complex plane: angular and radial dependence}

As the definition (\ref{rn0}) for $n=1$ indicates, the $Y$-dependence of $P_1$  manifests in  $R_1$ too.  The evolution equation for the latter  can then be obtained by  an integration of eq.(\ref{pcmp}) over all eigenvalues except one of them,  say $z$  (hereafter,  using notations $z_1$ and $z_2$  as  the real and imaginary parts  of $z$), 

\begin{eqnarray}
{\partial R_1(z)\over\partial Y} =  \sum_{s=1}^2 {\partial^2  R_1\over \partial z_s^2} +
 \sum_{s=1}^2 {\partial \over \partial z_s}\left( \gamma \, z_s \, R_1 - 2 \; {\bf P}\int {\rm d}^2 z' R_2 (z, z')  {z_s - z'_s\over |z- z'|^2}\right) 
 \label{r0z}
 \end{eqnarray}
with  ${\bf P}$ indicating  the principle part of the integral (details  in {\it appendix} C).   


A desirable next step would be to solve the above equation exactly but the presence of integral renders it technically difficult.  With eigenvalues distributed on the complex plane, a deeper insight in the $R_1$ behavior can be gained by analyzing  it in polar coordinates 
$r, \theta$. Substitution of  $z=r \; {\rm e}^{i \theta}, z' =r' \; {\rm e}^{i \theta'} $ reduces eq.(\ref{r0z}) as  (details  in {\it appendix} D)

\begin{eqnarray}
{\partial R_1(r, \theta)\over\partial Y}  =  \left[ {\partial^2  \over \partial r^2} + {1\over r^2} \; {\partial^2  \over \partial \theta^2} + 2 \gamma  + \left( \gamma  \, r + {2\over r} \right) \, {\partial \over \partial r} \right] \, R_1 + {1\over r} \, {\partial   ( r \, I_c)  \over \partial r} +  {1\over r} \,  {\partial I_d \over \partial \theta}  
\label{r1rt}
 \end{eqnarray}
where

\begin{eqnarray}
I_c(r, \theta) &=&  -\, 2 \, {\bf P}\int_0^{2\pi}   {\rm d}\theta' \, \int_0^{\infty} {\rm d} r'\,  C(r, \theta, r', \theta') \, R_2 ( r, \theta, r', \theta') 
 \label{ic}\\
I_d(r, \theta) &=& - \, 2 \, {\bf P} \int_0^{2\pi}   {\rm d}\theta' \, \int_0^{\infty} {\rm d} r'\ \, D(r, \theta, r', \theta') \, R_2 (r, \theta, r', \theta')
 \label{id} 
 \end{eqnarray}
and

\begin{eqnarray}
 C(r, \theta, r', \theta') &=& {r' (r - r' \; \cos(\theta - \theta')) \over r^2 + r'^2 - 2 r r' \cos(\theta -\theta') }, 
\label{ac} \\
  D(r, \theta, r', \theta')  &=& { r'^2 \, \sin(\theta - \theta') \over r^2 + r'^2 - 2 r r' \cos(\theta -\theta') }.
 \label{ad} 
 \end{eqnarray}
Hereafter  the  integration limits will not be mentioned explicitely  unless   needed for clarity.

 As the right side of eq.(\ref{r1rt}) is $Y$-independent, it can be solved by separation of variables approach. Considering a solution 

\begin{eqnarray}
 R_n(r, \theta; Y-Y_0)  = {\mathcal R}_n(r, \theta) \, {\rm e}^{- E \, (Y-Y_0)},
\label{r1sep} 
\end{eqnarray}
 with $n=1,2$ and $Y-Y_0 \ge 0$,  we have 

 \begin{eqnarray}
 \left[ {\partial^2  \over \partial r^2} + {1\over r^2} \; {\partial^2  \over \partial \theta^2} + 2 \gamma  + \left( \gamma  \, r + {2\over r} \right) \, {\partial \over \partial r} \right] \,  {\mathcal R}_1 + {1\over r} \, {\partial   ( r \, I_c)  \over \partial r} +  {1\over r} \,  {\partial I_d \over \partial \theta}  + E \,  {\mathcal R}_1 =0
\label{r1rt1}
 \end{eqnarray}
 with  $I_c$ and  $I_d$ again given by eq.(\ref{ic}) and eq.(\ref{id}) respectively but with $R_2$ replaced by ${\mathcal R}_2$.   Here $E \ge 0$ for finiteness of  $R_1$ as $Y-Y_0$ becomes large.

As both $I_c$ and $I_d$ are dependent on $r$ as well as  $\theta$, 
eq.(\ref{r1rt}) indicates a non-ergodic spectral density, correlated in $r$ and $\theta$ variables.  Based on certain approximations specific to $r$-range,  these  correlations  can however be neglected,  suggesting the separable solution 
\begin{eqnarray}
{\mathcal R}_1(r, \theta)= U(r) \, T(\theta)
\label{sep1}
\end{eqnarray}
As the approximations are different for different $r$-ranges,  we now consider the behavior of $R_1$ in three different $r$-regions and seek justification  for the above conjecture, both analytically and theoretically.

\subsection{\bf  $r \ll 1$, bulk region:} By expanding eq.(\ref{ac}) and eq.(\ref{ad}) in powers of $r$, we have $C(r, \theta, r', \theta') = -  \,\sum_{m=1}^{\infty}  \left({r\over r'}\right)^{m} \, \cos[m (\theta-\theta')]$ and $D(r, \theta, r', \theta') = \sum_{m=1}^{\infty}  \left({r\over r'}\right)^{m} \, \sin[m (\theta-\theta')]$.  Substitution of the above in eq.(\ref{ic}) and eq.(\ref{id}) leads to 

\begin{eqnarray}
I_c(r, \theta) &=&  2 \, \sum_{m=1}^{\infty} r^{m-1} \, {\tilde Q}_m(\theta) 
\label{icss1}  \\
I_d(r, \theta) &=&  - 2 \, \sum_{m=1}^{\infty} r^{m-1} \, {\tilde S}_m(\theta) \label{idss1}
 \end{eqnarray}
with 

\begin{eqnarray}
{\tilde Q}_m(r, \theta) &=& P \int  {\rm d}\theta' \, \cos[m (\theta-\theta')] \, \langle {\left(1\over r'\right)^m} \rangle_{\theta'} \label{qtm} \\
{\tilde S}_m(r, \theta) &=& P \int  {\rm d}\theta' \, \sin[m (\theta-\theta')] \, \langle {\left(1\over r'\right)^m} \rangle_{\theta'}   \label{stm} 
\end{eqnarray}
with $\langle {1\over r'^m} \rangle_{\theta'}=\int {{\rm d}r' \, r' \, {\mathcal R}_2\over r'^{m}}$. Using eq.(\ref{icss1}) and eq.(\ref{idss1}), we have
 
\begin{eqnarray}
{1\over r} \, {\partial   (r \, I_c)  \over \partial r} +  {1\over  r} \,  {\partial I_d \over \partial \theta}   = {2\over r} \, \sum_{m=1}^{\infty}  \left( {\partial (r^m \; {\tilde Q}_m) \over \partial r}  -  r^{m-1} \,
{\partial {\tilde S}_m\over \partial \theta}\right)
\label{icd1}
\end{eqnarray}
 As both $ {\tilde Q}_m$ and ${\tilde S}_m$ are $\theta$-dependent,   substitution of the above  leaves eq.(\ref{r1rt1}) non-separable in $r, \theta$ variables. In the regime $r \sim {1\over \sqrt{N}}$, however,  the series in eq.(\ref{icd1}) can contribute  significantly only if   $ {\tilde Q}_m$ and ${\tilde S}_m$ vary as $N^{(m-1)/2}$ or faster.

The latter not only requires $\langle \left({1\over r'}\right)^{m} \rangle_{\theta'} \sim N^{(m-1)/2}$  (such that $\int {{\rm d}r' \, {\mathcal R}_2\over r'^{m-1}} \sim N^{(m-1)/2}$) but also its  dependence on $\theta$ (to keep non-zero contributions from the integrals in eq.(\ref{icd1}) for all $m \ge 1$).  
But as $\lim_{r, r' \to 0} {\mathcal R}_2(r, r', \theta, \theta')$ is not very sensitive to angular dependence,  this in turn suggests a $\theta'$-independence of $\langle {1\over r'^m}\rangle_{\theta'} $ near origin ($r' \leq {1\over \sqrt{N}}$) and is also confirmed by  our numerical analysis of three different ensembles, discussed later in section V.  As a consequence, for almost all  cases (e.g except for those with phase singularities at the origin \cite{psps}),  the contribution of the series in eq.(\ref{icd1}) to eq.(\ref{r1rt1})  is negligible with respect to other terms and we can write

\begin{eqnarray}
   \left[ {\partial^2  \over \partial r^2} + {1\over r^2} \; {\partial^2  \over \partial \theta^2} +  {2\over r} \, {\partial \over \partial r} \right] \, {\mathcal R}_1 + E \, {\mathcal R}_1 =0
 \label{r1rts}
 \end{eqnarray}

Substituting eq.(\ref{sep1}) in eq.(\ref{r1rts}), we now have

\begin{eqnarray}
 r^2 \, {\partial^2 U\over \partial r^2} +   2  \, r \; {\partial U\over \partial r} + \left(E \, r^2 - E_1 \right)\, U &=&  0\,
 \label{us1} \\
{\partial^2 T( \theta) \over \partial \theta^2} + E_1 \, T  &=& 0
\label{ts1} 
\end{eqnarray}

Using separation of variables, the solution of above equations can be given as 

\begin{eqnarray}
 U_\mu(r) &=& r^{-1/2} \, \left[ a_{\mu}\, J_{\mu} \left({ r \sqrt{E}} \right) + b_{\mu} \, N_{\mu} \left(r \sqrt{E} \right) \right] 
 \label{us2} \\
 T_{\nu}( \theta, Y) &=& c_{\nu} \, \cos\left[\sqrt{E_1} \; \theta + d_{\nu} \right] 
\label{ts2} 
\end{eqnarray}
 with $J_{\mu}, N_{\mu}$ as the Bessel's functions, $E_1={1\over 4}- {\mu}^2$,  $a_{\mu}, b_{\mu}, c_{\nu}, d_{\nu}$ as arbitrary constants, determined by the boundary conditions on ${\mathcal R}_1(r, \theta)$; the condition that latter remains finite at $r=0$ then implies $b_{\mu}=0$.  The single valuedness of $T(\theta)$ as $\theta \to \theta + 2 \pi$ requires $d_{\nu}=2 \pi k$ and  $\sqrt{E_1}= n$ with $k, n$ as  integers and thereby $\mu =\pm \sqrt{1/4-n^2}$.  To keep $\mu$ as a real number, only allowed values of $n=0$.  The later in turn gives $\mu^2=1/4$.  Noting that $J_{-\mu} (x)= (-1)^n \, J_{\mu}(x)$, hereafter we use $\mu =1/2$.

As the arbitrary constant $E$ is not subjected to any additional known constraints except semi-positive definite one,   a valid particular solution can occur for continuous range of $E$ from $0 \to \infty$.  By substitution of the solutions in eq.(\ref{us2}) and eq.(\ref{ts2}) for $n=0$ (equivalently $E_1=0$) in eq.(\ref{r1sep}), the  general solution of eq.(\ref{r1rt}) for arbitrary initial condition (but finite at $r=0$) and for $0 \le r \le {1\over \sqrt{N}}$ can now be given as  

\begin{eqnarray}
R_1(r, \theta; Y) =   r^{-1/2} \, \int_0^{\infty}  {\rm d}E \; a(E) \, J_ {1/2}\left( r \sqrt{E} \right)  \, {\rm exp}\left[ - E\, (Y-Y_0)\right]   
\label{dsol}
\end{eqnarray}
with constant $a(E)$ determined by the initial condition of $R_1$ (examples discussed in {\it appendix} E).  A circular symmetric behavior of $R_1$ is also confirmed by our numerics discussed in section V, thus lending credence to our separability conjecture in eq.(\ref{sep1}).

The existence of a stationary solution in limit $Y-Y_0 \to \infty$ requires $E=0$. The solution in the limit  becomes  

\begin{eqnarray}
 R_1(r, \theta; \infty)=   \lim_{E\to 0} \;  {a(E) \over \sqrt{ r}}  \, J_ {1/2}\left(r \sqrt{E} \right)  ={1\over \sqrt{\pi}}  \hspace{0.3in} {\rm for} \; r \le {1\over \sqrt{N}}
\label{rinf1}
\end{eqnarray}
Here the  condition,that $\lim_{r \to 0} R_1(r, \theta)$ remains finite requires $a(E) \propto E^{-1/4}$; this in turn gives a constant $ R_1(r, \theta; \infty)$  for $r \le {1\over \sqrt{N}}$.

We note that the solution in eq.(\ref{dsol}) is obtained for smooth behavior of the density at $r=0$.  The solution for the cases in which initial spectral density is singular at $r=0$,  we must also consider  $b_{\mu} \not=0$.

 \vspace{0.2in}

\subsection{\bf $r \gg 1$, edge region:} Expanding eq.(\ref{ac}) and eq.(\ref{ad}) in powers of ${1\over r}$, we have $C(r, \theta, r', \theta')=  \sum_{m=0}^{\infty}  \left({r'\over r}\right)^{m+1} \, \cos[m (\theta-\theta')]$, 
$D(r, \theta, r', \theta') = \sum_{m=0}^{\infty}  \left({r'\over r}\right)^{m+1} \, \sin[m (\theta-\theta')]$.
Substitution of the above in eq.(\ref{ic}) and eq.(\ref{id}) leads to 

\begin{eqnarray}
I_c(r, \theta) &=&  - 2 \, \sum_{m=0}^{\infty} \left({1\over r}\right)^{m+1} \, Q_m(r, \theta) \label{icl1}  \\
I_d(r, \theta) &=& -  2  \, \sum_{m=0}^{\infty} \left({1\over r}\right)^{m+1}  \, S_m(r, \theta) \label{idl1}
 \end{eqnarray}
with 

\begin{eqnarray}
Q_m(r, \theta) &=& {\bf P} \int  {\rm d}\theta' \, \cos[m (\theta-\theta')] \; \langle { (r')^{m}}\rangle \label{qm}\\
S_m(r, \theta) &=& {\bf P} \int  {\rm d}\theta' \, \sin[m (\theta-\theta')] \; \langle {(r')^{m}}  \rangle. 
\label{sm}
\end{eqnarray}
with $\langle (r' )^m \rangle = \int {\rm d}r' \, r'^{m+1} \, {\mathcal R}_2(r, \theta, r', \theta')$ as a function of $r, \theta, \theta'$.
Using the above, we have

\begin{eqnarray}
{1\over r} \, {\partial   (r \, I_c)  \over \partial r} +  {1\over  r} \,  {\partial I_d \over \partial \theta}   = -{2\over r} \, \sum_{m=0}^{\infty} \, \left({\partial \over \partial r} \left(Q_m \over r^m\right)  + \left({1\over r}\right)^{m+1} 
{\partial S_m\over \partial \theta}\right)
\label{icd0}
\end{eqnarray}
 
Here   $Q_0(r, \theta) = \int  {\rm d}\theta' \, {\rm d}r' \, r' \, {\mathcal R}_2(r, \theta, r', \theta') = N \, {\mathcal R}_1(r, \theta)$ but $Q_m$ and $S_m$,  for $m >0$,  are complicated functions of $\theta$.   

Fortunately however the contribution from the terms containing $Q_m$, $S_{m+1}$ for $m >0$ can be neglected with respect to $Q_0$ and $S_1$ (with $S_0=0$) for an asymptotic decay in $R_1$ decay.  This can be explained as follows:
in the regime $r \sim \sqrt{N}$,  the series in eq.(\ref{icd0}) can contribute  significantly only if $Q_m$ and $S_m$  vary as $N^{m/2}$ or faster.  The latter in turn requires $\langle (r')^{m} \rangle_{\theta'} \sim N^{(m-1)/2}$ but this is not the case (as  $\int {\rm d}r' \,(r')^m {\mathcal R}_2$ is dominated by the behavior near origin due to decreasing spectral density for large $r$).  This is again confirmed by our numerical analysis of three different ensembles, discussed later in section V   Assuming this to be the case,  a substitution of the relation in eq.(\ref{r1rt}), retaining only  $m=0$ term,  now leads to 

 \begin{eqnarray}
\left[  {\partial^2  \over \partial r^2} + {1\over r^2} \; {\partial^2  \over \partial \theta^2} + 2 \gamma  + \left(\gamma \,r + {2\over  r} \right) \, {\partial \over \partial r} \right] \, {\mathcal R}_1 
 - {2\, N \over  r  }  \, \left( {\partial  {\mathcal R}_1  \over \partial r}   \right)  - {1\over r^2} {\partial S_1\over \partial \theta}  + E \; {\mathcal R}_1 =0
\label{r1rtl}
 \end{eqnarray}
 with $S_1(r, \theta)=  \int  {\rm d}\theta' \, {\rm d}r'  \sin(\theta-\theta')  \, r'^2 \, {\mathcal R}_2(r, \theta, r', \theta')$.

 As our interest here is in the cases  with  decaying spectral density away from the origin $r=0$ (i.e relatively lower spectral density in the edge region),  the contribution to $r'$-integral in $S_1$  is dominated by the bulk, this implies  $|r'-r|$ is large (with respect to mean level spacing $\sim N^{-1/2}$) for the dominant part of the $r'$-integral.  We can then safely assume ${\mathcal R}_2(r, \theta, r', \theta') \approx {\mathcal R}_1(r, \theta) \, {\mathcal R}_1(r', \theta')$.  (This is based on rewriting ${\mathcal R}_2$ in terms of the $2^{nd}$ order cluster function \cite{haak, circ,meta}: $T_2(r, \theta, r', \theta')= {\mathcal R}_2(r, \theta, r', \theta') - {\mathcal R}_1(r, \theta) \, {\mathcal R}_1(r', \theta')$. But as $T_2$ is of order $N^{-1}$ or lower)  smaller than the other term, we can approximate ${\mathcal R}_2(r, \theta, r', \theta') \approx  {\mathcal R}_1(r, \theta) \, {\mathcal R}_1(r', \theta')$.   The latter approximation is the standard route used in past for Hermitian as well as unitary ensembles (discussed in detail in section(6.14) and specifically eq.(6.14.5) of \cite{haak},  also discussed below eq.(59) in \cite{circ} in case of circular Brownian ensembles).  Using the above approximation we have $S_1(r, \theta)={\mathcal S}(\theta) \, {\mathcal R}_1(r, \theta)$ with 
 
\begin{eqnarray} 
 {\mathcal S}(\theta)=\int_0^{2\pi}  {\rm d}\theta' \, \int_0^{\infty} {\rm d}r'  \sin(\theta-\theta')  \, r'^2 \, {\mathcal R}_1(r', \theta').
 \label{stl}
 \end{eqnarray}
  
 Substitution of eq.(\ref{sep1}) now leads to (with  $S_0=0$) 
 
 \begin{eqnarray}
r^2 \, {\partial^2  U\over \partial r^2} + r \, (\gamma  \, r^2 +2 - 2 \, N)   {\partial U\over \partial r}   +  (2 \gamma + E) \, r^2 \, U  -  E_1  \, U  &= &  0 \label{ul1}\\
{\partial^2  T\over \partial \theta^2}  -  2 \, {\mathcal S} \,  {\partial  T \over \partial \theta}  + E_1 \, T  &=&  0
\label{tl1}
 \end{eqnarray}
 with $E_1$ as an arbitrary constant, determined by the boundary conditions that  ${\mathcal R}_1(r, \theta)$ vanishes at large radial distances and is single valued in $\theta$. This in turn corresponds to following constraints: $\lim_{r\to \infty} \, U(r) \to 0$ and $T(\theta; E_1) = T(2 \pi +\theta; E_1)$.   A particular solution of eq.(\ref{ul1}) for an arbitrary constant $E_u=2\gamma+E$ is 
 
\begin{eqnarray}
U(r, E, E_1) &=& \left({2\over \gamma \, r^2}\right)^{\frac{1-N_0}{4}}
\left( c \, \left({2\over \gamma \, r^2}\right)^{\frac{a}{4}} \, U_1 + d \, \left({2\over \gamma \, r^2}\right)^{-\frac{a}{4} } \, U_2 \right) 
\label{ui}
\end{eqnarray}
with  $U_1=F_1\left(\frac{E}{2 \gamma} +\frac{3+N_0 -a}{4}, \frac{2-a}{2} ,-\frac{\gamma r^2}{2}\right)$ and $U_2 =F_1\left[\frac{E}{2 \gamma}+\frac{3+N_0+a}{4}, \frac{2+a}{2},-\frac{\gamma r^2}{2} \right]$, 
$F_1$ as the confluent Hypergeometric function $ _1F_1(\alpha, \beta, x)$ \cite{temme}, 
$a=\sqrt{4 E_1+(N_0-1)^2}$, $N_0=2N$  and $c, d$ as arbitrary constants to be determined by the initial conditions. Noting that both $E_1, E$ appear along with $N_0$, their influence on the solution  can be seen only if they are of the order of $N_0$ or larger. We redefine them, without loss of generality (both constants being arbitrary), as $E_1= { (\mu^2-1) \over 4}\,  (N_0-1)^2$ and $E=2 \gamma \, \nu  \, N_0$ with $\mu, \nu $ as arbitrary constants  to be determined by boundary conditions; the conditions that $E,  E_1 \ge 0$ however also requires $\mu \ge 1$ and $\nu \ge 0$. This lead to $a=(N_0-1) \mu$ and thereby in large $N$ limit, 
$U_1 \approx F_1\left( \frac{N_0 (4 \nu-\mu+1)}{4},  -\frac{\mu N_0 }{2},  -\frac{\gamma r^2}{2}\right)$ 
 and $U_2 \approx F_1\left(\frac{N_0(4 \nu+\mu+1)}{4}, \frac{ \mu N_0}{2},-\frac{\gamma r^2}{2}\right)$.  The particular solution of eq.(\ref{ul1}) can now be given as 

\begin{eqnarray}
U_{\mu \nu}  &=&  c_{\mu \nu} \, \left({ \gamma \, r^2 \over 2}\right)^{-\frac{(N_0-1)(\mu-1)}{4}} \,   F_1\left(\frac{N_0(4 \nu-\mu+1)}{4}, \frac{ \mu N_0}{2},-\frac{\gamma r^2}{2}\right)  + \nonumber \\
& & d_{\mu \nu} \, \left({ \gamma \, r^2 \over 2}\right)^{\frac{(N_0-1)(\mu+1)}{4}} \,   F_1\left(\frac{N_0(4 \nu+\mu+1)}{4}, \frac{ \mu N_0}{2},-\frac{\gamma r^2}{2}\right)
\label{umnn}
\end{eqnarray}
As the above solution is valid in large $r$-limit,  the first term  with coefficient $c_{\mu \nu}$, for $\nu \not=0$,  is  negligible as compared to second term.   For $\nu=0$, however, the first term approaches a constant value.  The condition $\lim_{r \to \infty}{\mathcal R}_1(r, \theta) \to 0$ in eq.(\ref{ui}) therefore requires $c_{\mu 0}=0$.

The solution of eq.(\ref{tl1}) depends on the form of ${\mathcal S}$ in eq.(\ref{stl}). Using ${\mathcal R}_1(r, \theta) = U(r) \, T(\theta)$, we can rewrite ${\mathcal S}(\theta) = \langle r \rangle \, \left(a_c \, \sin(\theta) - a_s \, \cos(\theta) \right)$ with 
$a_c=\int_0^{2\pi}  {\rm d}\theta' \,   \cos(\theta')  \, T(\theta')$ and $a_s=\int_0^{2\pi} {\rm d}\theta' \,   \sin(\theta')  \, T(\theta')$.  For the case  with $a_c=0, a_s=0$,we have  ${\mathcal S}=0$.  Eq.(\ref{tl1}) then reduces to ${\partial^2  T\over \partial \theta^2} - E_1 \, T =  {\partial T\over\partial Y}$ and its solution can be given as that of eq.(\ref{ts1})).  For the case with $a_c \not= 0, a_s \not= 0$, we have ${\mathcal S}(\theta) \sim o(N)$ (with  $\langle r \rangle = \int_0^{\infty} {\rm d}r \, r^2 \, U(r))$ and $E_1=m q_1 N^2$; the second order derivative in eq.(\ref{tl1})  can then be neglected with respect to other terms.   For $a_c \not= 0, a_s \not= 0$,  the  solution of eq.(\ref{tl1})  can now be given as, 

\begin{eqnarray}
T(\theta) &=&  b\, {\rm exp}\left[ - {E_1 \over \langle r \rangle} \, I_0 \right]  
\label{tl2}
\end{eqnarray}
with $b$ as an arbitrary constant and

\begin{eqnarray}
I_0(\theta)=- {1\over 2}  \, \int {{\rm d}\theta \over  a_c \, \sin(\theta) - a_s \, \cos(\theta)}=  {1\over \sqrt{a_c^2 + a_s^2}}  \tan^{-1}\left[{a_c + a_s \tan(\theta/2) \over \sqrt{a_c^2 + a_s^2}}\right]
\label{ti0}
\end{eqnarray}

Eq.(\ref{tl2}) satisfies  the periodic condition on $T(\theta)$ as  $I_0(2 \pi +\theta)=I_0(\theta)$. Using $E_1={ (\mu^2-1) \over 4}\,  (2 N-1)^2$, the solution in eq.(\ref{tl2}) can be written as 

\begin{eqnarray}
T_{\mu}(\theta) =b_{\mu}  \, {\rm exp}\left[ - {(\mu^2-1) (2 N-1)^2 \over 4 \langle r \rangle} \, I_0(\theta) \right].
\label{tmn1}
\end{eqnarray}

By substitution of eq.(\ref{umn}), eq.(\ref{tl2}) in eq.(\ref{sep1}) and subsequently in eq.(\ref{r1sep}) for $n=1$, the general solution for $R_1(r, \theta)$ can now be given as

\begin{eqnarray}
R_1(r, \theta; Y) \approx  {1\over \langle r \rangle} \, \left({ \gamma \, r^2\over 2}\right)^{\frac{(2 N-1)}{4}} \, \sum_{\mu, \nu} \,  b_{\mu} \, U_{\mu \nu}(r)  \; \, {\rm e}^{ - {(\mu^2-1) (2 N-1)^2 \, I_0(\theta) \over 4 \langle r \rangle} } \, {\rm exp}\left[- 4 \gamma \, \nu  N  (Y-Y_0)\right] \nonumber \\
\label{ri} 
\end{eqnarray}
with  arbitrary constants $b_{\mu}$ and  $d_{\mu \nu} $ determined from the initial solution ${ R}_1(r, \theta; Y_0)$ (examples discussed in {\it appendix} E).

Here again for stationary solution to exist in limit $Y-Y_0 \to \infty$, we need $\nu=0$.  This leads to,  with $U_{\mu 0}$ given by eq.(\ref{umnn}), 

\begin{eqnarray}
R_1(r,\theta; \infty) \approx    {1\over \langle r \rangle} \,  \sum_{\mu} \, a_{\mu 0}   \, \left({ \gamma \, r^2\over 2}\right)^{\frac{(N_0-1)(\mu+1)}{4}}  \, F_1\left(\frac{N_0(\mu+1)}{4}, \frac{ \mu N_0}{2},-\frac{\gamma r^2}{2}\right) 
\, {\rm e}^{ -  {(\mu^2-1) \, (N_0-1)^2 \, I_0 \over 4 \langle r \rangle} } \nonumber \\
\label{ri4}
\end{eqnarray}
with $a_{\mu \nu}=b_{\mu}  \, d_{\mu \nu} $.
As the term $\mu=1$ is the dominant term in the above series, $R(r,\theta; \infty) $ can further be approximated as 

\begin{eqnarray}
R_1(r,\theta; \infty) \sim  {1\over \langle r \rangle}  \left({\gamma \, r^2 \over 2}\right)^{\frac{(N_0-1)}{2}} \,  {\rm e}^{-{\gamma r^2\over  2}} \hspace{0.5in} {\rm for} \;  r \sim \sqrt{N}  
\label{rinf2}
\end{eqnarray}
where we have used the identity $F_1(a,a,x)={\rm e}^x$ \cite{temme}.

A summation of series in eq.(\ref{ri}), leading to a closed form expression for $R_1$, depends on the initial condition  $R_1(r, \theta; Y_0)$ .  Nonetheless  eq.(\ref{ri}) gives an important insight even for arbitrary $a_{\mu \nu}$; while a summation over $\nu$ transforms $U_{\mu \nu}(r)$ to  a $Y$-dependent function,  the $\theta$-dependent part remains independent of $Y-Y_0$.  
 
\vspace{0.2in}

\subsection{\bf Arbitrary $r$ with rapidly decaying second order correlations:}
With both $C, D$ as well as $R_2$ dependent on four variables $r, \theta,  r', \theta'$,  the determination of $I_c$ and $I_d$ from eq.(\ref{ic}) and eq.(\ref{id}),  for arbitrary $r$, is technically complicated. Important insights in the solution can however be gained by a Taylor series expansion of $C$ and $D$ in the neighborhood of $r, \theta$.

\vspace{0.1in}

{\noindent  {\bf (i) Rapidly decaying angular moments of ${\mathcal R}_1(r, \theta)$  for arbitrary $r$:}}

\vspace{0.1in}

From eq.(\ref{ac}) and eq.(\ref{ad}), we note that ${\partial^{2n+1} C \over \partial \theta'^{2n+1}}=0,   \quad  {\partial^{2n} D \over \partial \theta'^{2n}}=0$      at $\theta=\theta'$ for $n$ as a positive integer and  ${\partial^{2n} C \over \partial \theta'^{2n}}$ and  ${\partial^{2n+1} D \over \partial \theta'^{2n+1}}$ at $\theta'=\theta$ have a singularity at $r'=r$.   $I_c,  I_d$ can then be simplified by a Taylor series expansion of $C$ and $D$ as powers of $S=\sin(\theta-\theta')$ in the neighborhood of $\theta=\theta'$.  While this results in a series of double integrals over $r', \theta'$,  the dominant contribution for each $r'$-integral comes from the region in neighborhood of $r=r'$, leading to 

\begin{eqnarray}
I_c(r, \theta) = - 2  \, \sum_{m=0}^{\infty} J_{2m}, \qquad
I_d(r, \theta)= -2 \, \sum_{m=0}^{\infty}  G_{2m+1}  
\label{icd6} 
\end{eqnarray}
with 

\begin{eqnarray}
J_{2m}(r, \theta) &=& \int {\rm d}r' \, u_{2m}(r, r') \, \langle S^{2m} \rangle_2  \\
G_{2m+1}(r, \theta) &=& \int {\rm d}r' \, {v_{2m+1}(r, r') \, \langle S^{2m+1}\rangle}_2
\label{icd7} 
\end{eqnarray}
where

\begin{eqnarray}
u_n(r,r')= {1 \over n!} \; {\partial^n C \over \partial S^n}\mid_{t=0},  \hspace{0.1in}
v_n(r,r')= {1 \over n!} \; {\partial^n D \over \partial S^n}\mid_{t=0}, \label{uvn}
\end{eqnarray}
and
$\langle S^n\rangle_2 = \int {\rm d} \theta' \, S^n \, {\mathcal R}_2(r, \theta, r', \theta')$ as the $n$th moment of ${\mathcal R}_2$ (equivalently that of {\it jpdf} $P$ of the eigenvalues) and is a function of variables $r,  r'$ and $\theta$ as well.   Some lower orders of $u_n, v_n$ can be given as follows:
$u_0(r, x)={x \over r-x}, u_1=0$, $u_2= {(r+x) x^2 \over 2 (x-r)^3}, u_3=0$,  
$u_4= {x^2 (r^3-5 r^2 x -5 r x^2+x^3)\over 8(r-x)^5}, u_5=0$, $u_{2m} \sim (r-x)^{-(2m+1)},  u_{2m+1}=0$, $v_1(r, x)={x^2\over (r-x)^2}, v_2=0$, $v_3= {r x^3 \over (r-x)^4}, v_4=0$,  
$v_5= {r x^3 (r^2-6 r x +x^2)\over 4(r-x)^6}, v_6=0$, $v_{2m+1} \sim (r-x)^{-2(2m+1)}, v_{2m+1}=0$.  Using the above, we have

\begin{eqnarray}
{1\over r} \, {\partial   (r \, I_c)  \over \partial r} +  {1\over  r} \,  {\partial I_d \over \partial \theta}   = -{2\over r} \, \sum_{m=0}^{\infty} \, \left({\partial  \left(r J_{2m} \right) \over \partial r}  + \left({1\over r}\right) 
{\partial G_{2m+1}\over \partial \theta}\right)
\label{icd5}
\end{eqnarray}
While the main contribution to $r'$-integral in both $J_{2m}$ and $G_{2m+1}$ comes only from the neighborhood of $r'=r$ (as both $u_{2m}$ and  $v_{2m+1}$ have a pole there),  the $\theta'$-integral is dominated by the region $|\theta'-\theta|=\pi/2$.  For ${\mathcal R}_2$, this implies  the correlation between two eigenvalues located at far-off points on the complex plane (although same radial distance but at an angle of $\pi/2$) and is expected to be negligible.  As ${\mathcal R}_2$ describes the second order spectral correlation at the scale of local spectral density $R_1$,  the correlation between such points is expected to be weak and  we can approximate ${\mathcal R}_2(r, \theta, r', \theta') \approx {\mathcal R}_1(r,\theta) \, {\mathcal R}_1(r', \theta')$.  This in turn gives
$\langle S^n\rangle_2  \approx  {\mathcal R}_1(r, \theta) \,  \langle S^n\rangle$ with  $\langle S^n \rangle \equiv \int {\rm d} \theta' \, S^n \, {\mathcal R}_1(r', \theta')$  and thereby 
$J_{2m}(r, \theta) ={\mathcal R}_1(r, \theta) \,   \int {\rm d}r' \, u_{2m}(r, r') \, \langle S^{2m} \rangle$, 
$G_{2m+1}(r, \theta) = {\mathcal R}_1(r, \theta) \,  \int {\rm d}r' \, {v_{2m+1}(r, r') \, \langle S^{2m+1}\rangle}$. Further noting that $-1 \le s \le 1$ and $0 \le {\mathcal R}_1 \le 1$,  it is reasonable to  assume that $\langle S^n\rangle$ become very small for $n >0$, thereby making the contribution  from $J_{2m}$ and $G_{2m+1}$ negligible relative to $J_0$.  This in turn permits eq.(\ref{icd5}) to be approximated as follows: ${1\over r} \, {\partial   (r \, I_c)  \over \partial r} +  {1\over  r} \,  {\partial I_d \over \partial \theta}   \approx -{2\over r} \, {\partial  \left(r J_0 \right) \over \partial r}  $  where $J_0(r, \theta) =\int {r' \, {\mathcal R}_2 \over r-r'}  \; {\rm d}r'  \, {\rm d}\theta'$.  Substitution of the  above approximations reduces eq.(\ref{r1rt}) to

 \begin{eqnarray}
\left[  {\partial^2  \over \partial r^2} + {1\over r^2} \; {\partial^2  \over \partial \theta^2} + 2 \gamma  + \left( \gamma \,  r + {2\over  r} \right) \, {\partial \over \partial r} \right] \, {\mathcal R}_1  + E \; {\mathcal R}_1 = {2 \over  r  }  \, {\partial  (r \, J_0 ) \over \partial r}   
\label{r1rtf}
 \end{eqnarray}
Due to,  $J_0$,  the above equation is still not separable in $r$ and $\theta$ variables and further progress can be made only by using case-specific forms of  ${\mathcal R_2}$. 

 As an example,  here  consider the case with weak $2nd$ order density correlations all over the complex plane that permits ${\mathcal R_2} = {\mathcal R_1}(r, \theta) . {\mathcal R_1}(r', \theta')$; (we note that the approximation is different from the one used above to calculate $J_{2m}$ and $G_{2m+1}$,  and is valid in general at distances $|r-r'| $ larger than local mean level spacing).  $J_0$ can then be expressed as 
 \begin{eqnarray}
 J_0 = f(r)  \; R_1(r, \theta),  \hspace{0.5in} f(r) =\int {r' \, {\mathcal R}_1(r', \theta') \over r-r'}  \; {\rm d}r'  \, {\rm d}\theta. 
 \label{j000}
 \end{eqnarray}
   Multiplying eq.(\ref{r1rt}) by $r^2$ and substituting ${\mathcal R}_1(r, \theta) = U(r) \; T(\theta)$, we now have 

\begin{eqnarray}
 r^2 \, {\partial^2 U\over \partial r^2} + r \, \left(\gamma  \, r^2 +2 -2 \, r \,  f \right) \; {\partial U\over \partial r} + (2 \gamma +E -2 f') \, r^2 \, U -2 r f U - E_1 \, U 
 &=& 0 \label{ut1} \\
{\partial^2 T( \theta) \over \partial \theta^2}  + E_1 \, T &=& 0
\label{ttt1} 
\end{eqnarray}
with $f' \equiv {\partial f\over \partial r}$,  $E_1$ as an arbitrary constant, determined by the boundary conditions that  ${\mathcal R}_1(r, \theta)$ vanishes at large radial distances and is single valued in $\theta$.  Further the first term in eq.(\ref{tt1}) can be neglected, due to its being $N$ times smaller than the other terms.  We note,  following from eq.(\ref{j000}),  the above equations are valid within the region on the complex plane where $R_1(r, \theta)$ is non-zero.

The solution of eq.(\ref{ut1}) depends on mathematical form of $f$ defined in eq.(\ref{j000}).  As the correlations depend on system conditions,  $f(r)$ can in general be of different form as $Y$ varies.  Here we consider three representative cases.  For the cases in which $f(r)={\alpha\over r}$ with $\alpha$ as a constant,  eq.(\ref{ut1}) is of the same  form as that of eq.(\ref{ul1}), except for $N$  in the latter is now replaced by $\alpha$.   The solution for $U(r)$ can then be given by eq.(\ref{ui}) but now  $N$ is replaced by $\alpha$ and $c\not=0$
(as $r$ is finite in the present case). Proceeding again as in previous case,  the solution now  becomes (with $E_1= {1\over 4} (\mu^2-1) \, (\alpha-1)^2$)

\begin{eqnarray}
U_{\mu \nu}(r) &=& \left({ \gamma \, r^2 \over 2}\right)^{\frac{2\alpha-1}{4}}
\left[ {\tilde p}_{\mu \nu} \, \left({ \gamma \, r^2 \over 2}\right)^{-\frac{(2\alpha-1) \mu}{4}} \; F_1\left(\frac{\alpha(4 \nu + 1 -\mu)}{2}, -\mu \, \alpha,  -\frac{\gamma r^2}{2}\right)  \right.\,  + \nonumber \\
&+ & {\tilde q}_{\mu \nu} \,  \left({ \gamma \, r^2 \over 2}\right)^{\frac{(2\alpha-1)\mu}{4}} \, \left. F_1\left(\frac{\alpha(4 \nu+1+\mu)}{2},  \mu \, \alpha,-\frac{\gamma r^2}{2}\right)
\right]
\label{uia1}
\end{eqnarray}
with arbitrary constants ${\tilde p}_{\mu \nu}, {\tilde q}_{\mu \nu}$ determined from initial condition at $Y=Y_0$.

Similarly for the cases in which eq.(\ref{j000}) gives $f(r)=\alpha \; r$  with $\alpha$ as a constant,  eq.(\ref{ut1}) is again of the same  form as that of eq.(\ref{ul1}), except for $\gamma$  in the latter is now replaced by $\gamma-2 \alpha$ and  $N$ by a zero.  Proceeding again as in previous case,  the solution now  becomes,  with $E_1= {1\over 4} (\mu^2-1)$ and $E=2 \gamma_1 \nu$,

\begin{eqnarray}
U_{\mu \nu}(r) &=& \left({ \gamma_1 \, r^2 \over 2}\right)^{-\frac{1}{4}}
\left[ {\overline p}_{\mu \nu} \, \left({ \gamma_1 \, r^2 \over 2}\right)^{-\frac{ \mu}{4}} \; F_1\left(\frac{(4 \nu + 3 -\mu)}{4}, \frac{2-\mu}{2},   -\frac{\gamma_1 r^2}{2}\right)  \right.\,  + \nonumber \\
&+ & {\overline q}_{\mu \nu} \,  \left({ \gamma \, r^2 \over 2}\right)^{\frac{\mu}{4}} \, \left. F_1\left(\frac{(4 \nu+3+\mu)}{4}, \frac{2+\mu}{2} ,-\frac{\gamma_1 r^2}{2}\right)
\right]
\label{uja1}
\end{eqnarray}
with $\gamma_1= \gamma -2 \alpha$ and ${\overline p},  {\overline q}$ as arbitrary constants.

Another case worth consideration is with $f(r)={\rm e}^{-\eta \, r}$.   Eq.(\ref{ut1}) is now approximately of the same  form (by neglecting terms containing $f$ and $f'$ with respect to constant terms and $r^2$)  as that of eq.(\ref{ul1}), except for $N$  in the latter is now replaced by a zero.   Further,  for technical simplification but without any loss of generality, we write $E$ as $E=  2 \gamma \nu \chi$,  with  $\nu$ again a non-negative integer  and $\chi$ as an arbitrarily small positive constant; the choice ensures  arbitrariness of $E$ as well as permits its almost continuous values.   (It is sufficient to choose $\chi \ll {1\over (Y-Y_0)}$ for  the present analysis). We now have 
\begin{eqnarray}
U_{\mu \nu}(r) &=& p_{\mu \nu} \, \left({ \gamma \, r^2 \over 2}\right)^{-\frac{(\mu+1)}{4}} \; F_1\left(\frac{4 \nu \chi + 3 - \mu}{4}, 1-\frac{\mu}{2} \,  -\frac{\gamma r^2}{2}\right)    + \nonumber \\
&+ & q_{\mu \nu} \,  \left({ \gamma \, r^2 \over 2}\right)^{\frac{(\mu-1)}{4}} \, \left. F_1\left(\frac{4 \nu \chi + 3+\mu}{4},  1+ \frac{\mu}{2},-\frac{\gamma r^2}{2}\right) \right.
\label{uka1}
\end{eqnarray}

Contrary to radial evolution, the angular evolution described by eq.(\ref{ttt1})  differs from eq.(\ref{tl1}) and is similar to eq.(\ref{ts1}); its solution can then be given by eq.(\ref{ts2}); here again singlevaluedness of the solution requires $d_{\nu}=0$ but now $E_1= {1\over 4} (\mu^2-1) \, \alpha_1^2$ where $\alpha_1=(\alpha-1)$ for the cases $f=\alpha/ r$ and $f= \alpha r$;   $\alpha_1=1 $ for  the case $f={\rm e}^{-\eta r}$.

By substitution of appropriate $U_{\mu \nu}$ along with eq.(\ref{ts2}) in eq.(\ref{sep1}) and subsequently in eq.(\ref{r1sep}) for $n=1$, the general solution for $R_1(r, \theta)$ can  now be given as 

\begin{eqnarray}
R_1(r, \theta; Y) &\approx &   \sum_{\mu, \nu} \,  b_{\mu} \, U_{\mu \nu}(r) \;  \cos\left(\frac{\alpha_1}{2} \, \sqrt{\mu^2-1} \,  \, \theta \right)  \, {\rm e}^{- E  (Y-Y_0) }  
\label{rij} 
\end{eqnarray}
where $E= 4 \gamma \, \nu  \alpha $ for $f=\alpha/r$ and $f=\alpha \, r$, 
$E= 2 \gamma \nu \chi$ for $f={\rm e}^{-\eta r}$ and the arbitrary constants  in the above solution determined from the initial solution ${ R}_1(r, \theta; Y_0)$ (examples discussed in {\it appendix} E). 
Here again  a summation over $\nu$ transforms $U_{\mu \nu}(r)$ to $Y$-dependent function, leaving $\theta$-dependent part independent of $Y$.   The latter implies  $\theta$-dependence of $R_1(r, \theta; Y)$ does not vary significantly with $Y$.  As discussed in section V, this is confirmed by our numerical analysis too which again supports our conjecture in eq.(\ref{sep1}) (as the result in eq.(\ref{rij}) depends on the conjecture). Using  eq.(\ref{rij}),    $R_1(r, \theta; \infty)$ in this case again corresponds to  $E=0$ (equivalently  $\nu=0$)  and can be given as 

\begin{eqnarray}
R_1(r,\theta; \infty) \approx    \sum_{\mu} \,  b_{\mu} \,  U_{\mu 0} \;
  \cos\left(\frac{\alpha_1}{2} \, \sqrt{\mu^2-1} \, \theta \right) 
\label{rinf3}
\end{eqnarray}
Here again retaining only the dominant term corresponding to $\mu=1$ gives $R(r,\theta; \infty) \approx  b_1 \, U_{10}(r)$.   From eq.(\ref{uia1}) and eq.(\ref{uja1}),  we then have 

\begin{eqnarray}
R_1(r,\theta; \infty) &\approx&    {\tilde p}_{10}  +  {\tilde q}_{10} \,  \left({ \gamma \, r^2 \over 2}\right)^{\frac{(2\alpha-1)}{2}} \, {\rm e}^{-\frac{\gamma r^2}{2}}  
\hspace{1.7in} f={\alpha \over r}\\
&\approx&   p_{10} \, \left({ \gamma_1 \, r^2 \over 2}\right)^{-\frac{ 1}{2}} \; {\rm e}^{-\frac{\gamma_1 r^2}{2}} +  q_{10} \;  F_1\left(1, \frac{3}{2} ,-\frac{\gamma_1 r^2}{2}\right) \hspace{0.5in} f={\rm e}^{-\eta r}
\label{rinff}
\end{eqnarray}
Here we have used  $F_1\left(0, - \alpha,  -\frac{\gamma r^2}{2}\right) =1$ and $F_1\left(\alpha,   \alpha,-\frac{\gamma r^2}{2}\right)={\rm e}^{-\frac{\gamma r^2}{2}}$.  The above  indicates a constant density within the bulk spectral region for  the case $f={\alpha \over r}$.  Further the $Y \to \infty$ solution  for $f={\alpha \, r}$ case is of same form as in eq.(\ref{rinff}) but with $p_{10}, q_{10}$ replaced by $\overline{p}_{10},  \overline{q}_{10}$. 
We also note that, for a finite $R_1(r,\theta; \infty) $ at $r=0$ in for  the case $f={\alpha \, r}$ or  ${\rm e}^{-\eta r}$,  we must have  $\overline{p}_{10}=0$ and $p_{10}=0$.

We recall that the solutions (\ref{rij}) and (\ref{rinf3}) are applicable for the case with weak second order spectral correlations in the complex plane leading to $f(r)=\frac{\alpha}{ r},  \alpha \, r$ or $ {\rm e}^{-\eta r}$.   We note that the solution for the last case remains valid for  $f(r)$ decalying  faster than expoenential  too.   The solutions for other cases can be obtained by case specific approximation of eq.(\ref{r1rtf}).

\vspace{0.5in}

{\noindent  {\bf (ii) Slowly varying radial moments of ${\mathcal R}_1(r, \theta)$ around arbitrary $r, \theta$:}}

\vspace{0.1in}

For such cases, we can expand $C$ and $D$ in the neighborhood of $r'=r$. This leads to 

\begin{eqnarray}
 C(r, \theta, r', \theta') = r' \sum_{m=1}^{\infty}  {(r'-r)^{m-1}\over (2 r)^m} \, f_m (t)
 \label{aac1}
 \end{eqnarray}
 where $f_1(t)=1, \, f_2(t)={2\over t-1}, \, f_3(t)={2\over t-1}, \, f_4(t)={4 t\over (t-1)^2}, \, f_5(t)=-{4 (2t+1)\over (t-1)^2}$ with $t=\cos(\theta-\theta')$.
Substitution of eq.(\ref{aac1}) in eq.(\ref{ic}) gives (details in appendix)

\begin{eqnarray}
I_c(r, \theta) &=& -{N \over r} \, {\mathcal R}_1(r, \theta) -2 \, \sum_{m=2}^{\infty} {K_m \over (2 r)^m}.
 \label{ic1}
 \end{eqnarray}
Here the first term in the series is obtained by using $\int  {\rm d}\theta'  \, {\rm d} r' \, r' \, R_2 (r, \theta, r', \theta') = N \, {\mathcal R}_1(r, \theta)$ and 

\begin{eqnarray}
K_m(r, \theta) =  \int  {\rm d}\theta' \,  f_m (\cos(\theta-\theta'))  \; \mu_m(r, \theta, \theta')
\label{km}
\end{eqnarray}
with $\mu_m(r, \theta, \theta')$ as the $m^{th}$  order moment of $R_2 (r, \theta,  r', \theta') $ about a point $r'=r$ and for a fixed $r, \theta$ and $\theta'$ and defined as $\mu_m =  \int  {\rm d} r'  \,r' \,   (r'-r)^{m}\, {\mathcal R}_2 (r, \theta,  r', \theta')$.

The integral $I_d$ in eq.(\ref{id}) can similarly be approximated by  Taylor series  expansion of eq.(\ref{ad}); the latter gives

\begin{eqnarray}
 D(r, \theta, r', \theta') = {r'\over 2 \, r } \; g_0 (\cos(\theta-\theta')) + {r'\over r } \sum_{m=1}^{\infty}  {(r'-r)^{m}\over (2 r)^m} \, g_m (\cos(\theta-\theta'))
 \label{aa2}
 \end{eqnarray}
 where $g_0(t)=g_1(t)={v\over  (t-1)}, \, g_2(t)={v\over (t-1)^2}, \, g_3(t)={2 v \over (t-1)^3}, \, g_4(t)={-4 v \over (t-1)^3}$ etc with $v=\sin(\theta-\theta')$ and $t=\cos(\theta-\theta')$ (obtained by {\it Mathematica}). Substitution of eq.(\ref{aa2}) in eq.(\ref{id}) gives
 
\begin{eqnarray}
I_d(r, \theta) &=& - {L_0\over r} - {2\over r} \sum_{m=1}^{\infty} {L_m \over (2 r)^m} 
\label{id1}
\end{eqnarray}
 where  
 
\begin{eqnarray}
L_m(r, \theta) &=&   \int {\rm d} \theta'  \, g_m (\cos(\theta-\theta')) \, \mu_m(r, \theta, \theta')
\label{lm}
\end{eqnarray}

The series expansion of $I_c, I_d$ in eqs.(\ref{ic1}, \ref{id1}) are exact but their substitution in eq.(\ref{r1rt}) again leads to a  differential equation non-separable in $r, \theta$ variable and not easy to solve.  As clear from eq.(\ref{km}) and eq.(\ref{lm}), the non-separability arises from the terms with $K_m$ and $L_m$,  $m \ge 1$, in the series.  Fortunately however the contribution from $K_m$ to eq.(\ref{ic1})  is negligible for the spectral distributions with slowly varying moments around point $r, \theta$; this is because that the dominant contribution  from  the integration over $\theta'$ comes from the neighborhood of $t=1$ i.e equivalently $\theta'=\theta$.   Assuming $\mu(r, \theta, \theta')$ as a slowly varying function of $\theta'$,  it can then be  approximated  by its value at $\theta'=\theta$. We then have 

\begin{eqnarray}
K_m(r, \theta) \approx \mu_m(r, \theta, \theta) \;  \int  {\rm d}\theta' \,  f_m (\cos(\theta-\theta')) = 0   \qquad m >1
\label{km1}
\end{eqnarray}
The above equality follows  due to $ \int_0^{2\pi}  {\rm d}\theta' \,  f_m (\cos(\theta-\theta')) = 0$.  
Although, the $\theta'$-integration   in case of $L_m$ is not dominated by the region  around $t=1$,  we have  $L_m \ll L_1$.  Thus  if the moments $\mu_m(r, \theta, \theta') = \langle (r'-r)^{m} \rangle$ are slowly varying functions of $\theta'$ around arbitrary $r$,    the contribution from the higher order terms,  i.e those with $m >1$ in eq.(\ref{ic1}) and  with $m \ge 1$ in eq.(\ref{id1})),  can be neglected as compared to  the first term. This in turn permits following approximations

\begin{eqnarray}
I_c \approx -{N\over r} \, {\mathcal R}_1(r, \theta), \hspace{0.1in} I_d \approx - {L_0\over r} 
\label{icd}
\end{eqnarray}
Here the first term in $ I_c$ is obtained by using  the relation $\int  {\rm d}\theta'  \, {\rm d} r' \, r' \, R_2 (r, \theta, r', \theta') = N \, {\mathcal R}_1(r, \theta)$.

Substitution of the approximations (\ref{icd})  reduces eq.(\ref{r1rt}) separable in $r$ and $\theta$ variables. This can be seen as follows.  Multiplying eq.(\ref{r1rt}) by $r^2$ and substituting ${\mathcal R}_1(r, \theta) = U(r) \; T(\theta)$, we now have 

\begin{eqnarray}
 r^2 \, {\partial^2 U\over \partial r^2} + r \, \left(\gamma  \, r^2 +2 -N \right) \; {\partial U\over \partial r} + (2 \gamma +E) \, r^2 \, U - E_1 \, U 
 &=& 0 \label{u1} \\
{\partial^2 T( \theta) \over \partial \theta^2}  - N  {\partial \over \partial \theta}  \int  {\rm d}\theta'   \;   { \cot \left({\theta-\theta' \over 2}\right)}  \; T_2 (\theta, \theta') + E_1 \, T &=& 0
\label{tt1} 
\end{eqnarray}
with $E_1$ as an arbitrary constant, determined by the boundary conditions that  ${\mathcal R}_1(r, \theta)$ vanishes at large radial distances and is single valued in $\theta$.  Further the first term in eq.(\ref{tt1}) can be neglected, due to its being $N$ times smaller than the other terms.

As in previous case, here again eq.(\ref{u1}) is same as eq.(\ref{ul1}) except for $N_0=N$ in the former and $N_0=2 N$ in the latter; the solution for $U(r)$ can then again be given by eq.(\ref{ui}), using $N_0=N$ and $r$ finite.  This again leads to eq.(\ref{uia1})  but with $E=2 \gamma  \, \nu \, N$ and $E_1=(\mu^2-1)(N-1)^2$. 

Contrary to radial evolution, the angular evolution described by eq.(\ref{tt1})  differs from eq.(\ref{tl1}) and is seemingly more complicated. Important insight in its solution  can however be obtained by noting following analogy: substitution of the relation 

\begin{eqnarray}
{\overline\rho}(\theta; Y) = {\rm e}^{-E_1 (Y-Y_0)/N} \, T(\theta)
\label{circ1}
\end{eqnarray}
reduces  eq.(59) of \cite{circ}  in the same form as that of eq.(\ref{tt1}).  

Eq.(59) of \cite{circ}  however  describes  the evolution of the ensemble averaged spectral density ${\overline\rho}$ of the circular Brownian ensembles of unitary matrices, with eigenvalues distributed on a unit circle) in terms of a perturbation parameter $Y$ \cite{circ} for arbitrary initial conditions; (note here too the second order derivative was neglected due to negligible contribution in large $N$-limit)
Its general  solution is given as ${\overline\rho}(\theta; Y) = \sum_{E_1} {\rm e}^{-E_1 (Y-Y_0)/N} \, T(\theta; E_1)$ with permitted values of the constant $E_1$ given by the boundary conditions.   The above implies that $T(\theta)$ of our case corresponds to a particular solution of the average spectral density of the circular Brownian ensemble.

As for large $Y$, the evolution  in circular ensemble case is known to  approach uniform distribution with ${\overline\rho}(\theta; Y)  \approx {N\over 2 \pi}$ irrespective of the initial conditions.  This in turn suggest  $T(\theta,E_1) \approx {N\over 2\pi}$.  
Further if the initial condition at $Y=Y_0$ is a uniform density e.g  ${\overline\rho}(\theta,0)={N\over 2 \pi}$, the same behavior then persists for all $Y$; this again implies  $T(\theta,E_1)={N\over 2\pi}$. We note a uniform distribution of $T(\theta)$ is also indicated by our numerics of three different Gaussian ensembles of complex matrices   discussed in next section.

Substitution of eq.(\ref{uia1})  in eq.(\ref{r1sep}) along with $T(\theta, E_1)  \approx  {N \over 2 \pi}$,  the general solution for $R_1(r, \theta)$ in the present case can  be given as

\begin{eqnarray}
R_1(r, \theta; Y) \approx  {N \over 2 \pi} \sum_{\mu, \nu} \,  a_{\mu \nu} \,  U_{\mu \nu}(r) \; {\rm e}^{- 2 \gamma \, \nu  N (Y-Y_0)}
\label{rir} 
\end{eqnarray}
with the arbitrary constants $a_{\mu \nu}$  again  determined from the initial solution ${ R}_1(r, \theta; Y_0)$.  Again,  with condition $\nu=0$ in $Y-Y_0 \to \infty$ limit,  we have  $R_1(r,\theta; \infty) \approx   {N \over 2 \pi} \sum_{\mu} \, a_{\mu 0}  \, U_{\mu 0}(r)   $.

\subsection{\bf Comparison with Ginibre ensemble:}  

As mentioned in section II, the evolution of  the ensemble density (\ref{rho1})  approaches, in the limit $Y \to \infty$,  an equilibrium steady state described by Ginibre ensemble; thus corresponding ${\mathcal R}_1$  must also approach that of Ginibre ensemble in the above limit.   A previous study \cite{haak} (in section 8.8.2 therein) gives $R_{1, Ginibre} = {N \over \pi} \, {\rm e}^{- r^2} \sum_{k=0}^N {r^{2k} \over k!}$ for finite $N$; It approches a uniform distribution within a radius $\sqrt{N}$ in large $N$ limit: $R_{1, Ginibre} = {1\over \pi}$.  It is instructive to compare  the above result with  $Y \to \infty$ limit of our results.  From eqs.(\ref{rinf1},  \ref{rinf2}, \ref{rinf3}, \ref{rinf4}), we have $R_1(r, \theta, \infty) \to {\rm constant}$ within $r \sim \sqrt{N}$ and decaying exponentially beyond the regime and is a constant in $\theta$ too.    Thus our results are  consistent with \cite{haak} in large $N$ limit;

\section{Numerical analysis of spectral correlations}

Our primary  focus in this study  is to numerically analyze following two aspects of the spectral density: (i) the non-ergodic as well as non stationary aspect i.e dependence  on the energy range,  (ii)  the system dependence.

Our first objective has its roots  in the immense interest generated by non-ergodic aspects of many body spectrum for Hermitian as well non-Hermitian operators; the ergodicity, or its absence, is characteristic of a system's approach to thermalization which in turn facilitates the application of standard statistical tools. The motivation for the second objective comes from the unavoidable presence of disorder as well as many body interactions in physical systems which can manifests in a variety of ways in the matrix representation of their non-Hermitian operators; it is natural to query whether presence of disorder with/ without  interactions  has any, and if so, what impact on the average spectral density.

To achieve the objectives mentioned above, here we numerically analyze three  ensembles of non-hermitian random matrices with independent, Gaussian distributed  matrix elements  with zero mean but different functional dependence of their variances.  The ensemble density $\rho(H)$  in each case is given by eq.(\ref{pdf}), with $x_{kl;s}=0$ for all $k,l$ and $s$ indices but  $y_{kl;s}$ varying among elements. The details of the latter for the three ensembles are as follows: 

(i) {\it Ensembles with same off-diagonal to diagonal variance for all matrix elements, referred here as Brownian ensemble (BE)}: $y_{kl;s}= {1\over 2} ({1+N/b})^2, \quad y_{kk;s}={1\over 2} $. From eq.(\ref{yt1b}), this corresponds to (for a large $\gamma$),
\begin{eqnarray}
Y \approx    - 2 \; {\rm ln} \; (1 + N/b)   + C_0. 
\label{ybe}
\end{eqnarray} 
with $C_0$ as a constant. Choosing  initial BE with $b=b_0$ then gives $Y-Y_0=  - 2  \; {\rm ln} \;{ (1 + N/b) \over (1+N/b_0)}$.

 We recall  that our  theoretical formulation of the spectral density in section IV is based on standard assumption of $Y-Y_0 \ge 0$ (as mentioned below eq.(\ref{r1sep}).   For comparison with  our numerical analysis,  it is therefore appropriate to  ensure $Y-Y_0 \ge 0$   with increasing $b > b_0$.  For this reason, eq.(\ref{ybe})  is obtained by choosing negative sign in eq.(\ref{yt1b}).  The same reasoning will also be used for eq.(\ref{ype}) and eq.(\ref{yee}) given below.

(ii) {\it Ensembles with  off-diagonal to diagonal variance decaying as a power law, referred here as just power law ensemble  (PE) for brevity}: $y_{kl;s}={1\over 2} \left( 1+ (|i-j|/b)^2 \right)^2, \quad y_{kk;s}={1\over 2} $. Eq.(\ref{yt1b})
now gives 

\begin{eqnarray}
Y =  - {2\over 2 N^2} \sum_ {r=0}^N  g_{r} \, (N-r) \, {\rm ln}(1 + (r/b)^2)  + C_0. 
\label{ype}
\end{eqnarray}  
with $g_r=2-\delta_{r0}$.
Taking initial PE with $b=b_0$ then gives $Y-Y_0= - {2 \over N^2} \sum_ {r=1}^N \, (N-r) \, {\rm ln} \; {(1 + (r/b)^2) \over (1 + (r/b_0)^2)}$.

(iii) {\it Ensembles with  off-diagonal to diagonal variance decaying as an exponential, referred here as exponential  ensemble  EE}: $y_{kl;s}= {1\over 2} \, {\rm e}^{|k-l|^2/b^2} ,  \quad y_{kk;s}={1\over 2} $. Eq.(\ref{yt1b})
now gives 

\begin{eqnarray}
Y =  - {1\over 2 N^2} \sum_ {r=0}^N g_{r} \, (N-r) \, {\rm ln} \; ({\rm e}^{r^2/b^2} )   + C_0. 
\label{yee}
\end{eqnarray}  
with $g_r$ same as in eq.(\ref{ype}).
Taking initial EE with $b=b_0$ then gives $Y-Y_0= - {1\over N^2} \sum_ {r=1}^N  \, (N-r) \, \left({r^2\over b^2}- {r^2\over b_0^2} \right)$.

As indicated above,  while each ensemble depends on at least two system parameters, namely, $b$ and system size $N$, the nature of that dependence varies significantly, from constant behavior to  power law decay to exponential decay of the off-diagonals.  More clearly the off-diagonals in the PE and EE case are also functions of the distance between basis-states,  and thereby dependent on basis parameters.  To understand the system dependence of the spectral density,  we exactly diagonalize  each ensemble for many $b$ values and for matrix size $N= 1024$.  The ensemble size ${\mathcal M}$  i.e the number of  matrices in each ensemble is chosen  so as to give  a smooth behavior and therefore depends on the specific measure.  While the limit $b \to \infty$ in each case corresponds to  the Ginibre ensemble limit, the limit $b \to 0$ the Poisson statistics;  a variation of $b$ therefore leads to a Poisson $\to$ Ginibre crossover for each ensemble.  
Based on our theoretical prediction,  the Ginibre limit is reached when almost all off-diagonals of a typical matrix in the ensemble  become of the same order as the diagonals.  As this occurs at different rates for BE, PE and EE,  their  $b$-values for the Ginibre limit are different;  (this is indeed due to role of hidden system parameters manifesting through different variance structure for each ensemble).   With increasing $b$,  $y_{kl;s}$ for each of the three cases mentioned above approach almost same value,   with matrix elements distribution becoming identical.  The eigenvalues are then expected to increasingly repel each other with increasing $b$, thereby causing them to distribute uniformly on the complex plane.  and approach the Ginibre limit.  This is indeed confirmed by the figures 1-3,  illustrating the variation of spectral density on the $(r, \theta)$-plane as $b$ varies for BE, PE and EE respectively; here left panel depicts the behavior for a single matrix and right for ${\mathcal M}$.   The rate of change with $b$ however is different for each case e.g.  the approach to Ginibre limit is most rapid for EE case.

We recall that the Ginibre limit  corresponds to the ergodic limit.   It is therefore natural  to seek whether the spectral density  for each ensemble is indeed non-ergodic for smaller $b$-values?  To  illustrate the latter,  figures 1-3 display a comparison of the  spectral and ensemble averaging for the  spectral density  in each ensemble for four $b$-values. The spectral averaging  here is obtained by considering the density  of the eigenvalues of a single matrix on the $(r, \theta)$-plane  and the ensemble averaging  by averaging over the spectral density of the eigenvalues for three matrices.  As the illustrations in figures 1-3 indicate,  $\rho_{sm}(r, \theta)$ even for a single  matrix for large $b$ values ($b \sim N$),  appears almost same as $\langle \rho_{sm}(r, \theta)$ as well as $R_1(r, \theta)/N$ for all point on the complex plane reconfirming its ergodicity.  As $b$ decreases, a deviation between the two cases is noticeable even visually; this indicates a non ergodic nature of the spectrum  for small $b$ values in each case.   However the approach of the spectral density to ergodic limit as $b$ increases is different for each ensemble.

While the illustrations in figures 1-3 provide a qualitative insight  about the behavior of spectral density on the complex plane.,  it is desirable to display the density in a way that could give  better quantitative insights e.g.  by analyzing its radial/ angular dependence separately.  We also recall that the results obtained by our theoretical analysis are based on a separation of variables approach  i.e by assuming $R_1(r, \theta; Y)={\mathcal R}_1(r, \theta) \; {\rm e}^{-E(Y-Y_0)}$ along with ${\mathcal R}_1(r, \theta)= U(r) \; T(\theta)$ as a particular solution of the evolution equation for $R_1$.     This in turn implies that $\langle {\mathcal R}_1 \rangle_{\theta}  \equiv \int {\mathcal R}_1(r, \theta) \; {\rm d}\theta = constant .\; U(r) $ and therefore a numerical  averaging of ${\mathcal R}_1$ over $\theta$ is expected to agree with our theoretical result for $U(r)$  if the separability assumption is valid.  Similarly  $\langle {\mathcal R}_1 \rangle_{r}  \equiv \int {\mathcal R}_1(r, \theta) \; r \; {\rm d}r = constant .  T(\theta) $ and  a numerical  average of ${\mathcal R}_1$ over $r$ is also expected to agree with our theoretical result for $T(\theta)$.  As discussed in  section IV. A, B, C as well as in {\it appendix F},    $\langle R_1 \rangle_{r}$ is  almost constant in $\theta$ but $\langle R_1 \rangle_{\theta}$  varies with both $r$ and $Y$.

As mentioned above,  the Poisson $\to$ Ginibre crossover occurs at different rates for BE, PE and EE; it is therefore  natural to seek the detailed  role of $Y$ in  the crossover.   For numerical analysis,  we choose the  initial
condition  $b_0=1/N$ for each ensemble mentioned above.; the choice is theoretically predicted to  correspond to Poisson spectral statistics. (We note that the choice is arbitrary and one could equally well choose $b=0$ too).  
With Poisson ensemble as an initial condition,  
the approximation ${\mathcal R}_2(r,\theta, r', \theta') \approx {\mathcal R}_1(r,\theta) {\mathcal R}_1 (r', \theta')$  (used in section IV.C) is indeed valid for $Y=Y_0$ and also for small $Y-Y_0$, thereby justifying use of eq.(\ref{j000}).  
The numerics for  $b_0=1/N$ gives the initial density as 

\begin{eqnarray}
R_1(r, \theta; Y_0) \equiv R_1(r; Y_0) = A \; r^{-1/2} \, J_{1/2}(B r)  \; {\rm e}^{-C r^2}.
\label{r1th}
\end{eqnarray}
with $A, B, C$  as constants (the values  for each ensemble given in table I).  Using the above initial condition then suggest that $f(r)$ (eq.(\ref{j000})) decays exponentially or faster with $r$.  Thus  using eq.(\ref{rio3a}) along with  $U_{1 \nu}$ given by eq.(\ref{uka1}) and $\phi= 2 \gamma \chi (Y-Y_0)$,  we have 

\begin{eqnarray}
 \langle R_1(r,\theta,Y) \rangle_{\theta}  &=&
 \sum_{\nu=0}^{\infty}  \; U_{1\nu} \; {\rm e}^{-\nu \phi}  \label{rim2x} \\
&&\approx   \langle R_1(r,\theta,Y_0) \rangle_{\theta}  -   \phi \, \sum_{ \nu=1}^{\infty}  \; \nu \; U_{1\nu}
\label{rim2} 
\end{eqnarray}
with $U_{1 \nu} \equiv \left[ p_{1 \nu} \, \left({ \gamma \, r^2 \over 2}\right)^{-\frac{1}{2}} \; F_1\left(\frac{1}{2}+\nu \chi,  \frac{1}{2}, \,  -\frac{\gamma r^2}{2}\right)  + q_{1 \nu} \, \, F_1\left( \nu \chi +1,  \frac{3}{2},-\frac{\gamma r^2}{2}\right)  \right] $.
Here,  to keep  $\langle R_1(r,\theta,Y) \rangle_{\theta}$ finite at $r=0$,  we  need $p_{1\nu}=0 $;  (this follows from the relation $F_1\left(\frac{1+\nu \chi}{2},  \frac{1}{2} \,  -\frac{\gamma r^2}{2}\right) \approx {\rm e}^{-\frac{\gamma r^2}{2}}$. valid for small $\chi$).   As clear from the above,   $\langle R_1 (r, \theta,Y)\rangle_{\theta} $ for small $(Y-Y_0)$ remains almost of the same form as the initial density except for a change of coefficients.

Further for  large $(Y-Y_0)$, the density is expected to reach uniformity on the complex plane within a circle of radius $\sqrt{N}$;  the appropriate $U_{\mu \nu}$ in this case is therefore given by eq.(\ref{uia1}).   Using the latter in eq.(\ref{rima}), we have (now $\phi = 4 \gamma \alpha (Y-Y_0)$)

\begin{eqnarray}
&& \langle R_1 (r, \theta,Y)\rangle_{\theta}  \approx  
   \langle R_1(r,  \theta; \infty) \rangle_{\theta}  \; + \nonumber \\
& &\left({\tilde p}_{11} \, F_1\left( \alpha, -\alpha,  \frac{\gamma r^2}{2}\right) + 
 {\tilde q}_{11}  \; \left({ \gamma \, r^2 \over 2}\right)^{\frac{(2\alpha-1)}{2}} \;  F_1\left(- 2 \alpha,  \alpha,  \frac{\gamma r^2}{2}\right) \right) \ {\rm e}^{-\phi- \gamma \, r^2/2}  
\label{rim1} 
\end{eqnarray}
with  $ \langle R_1(r,  \theta; \infty) \rangle_{\theta} =\left({\tilde p}_{1 0}  + {\tilde q}_{10} \; \left({ \gamma \, r^2 \over 2}\right)^{\frac{(2\alpha-1)}{2}} \;  {\rm e}^{-\gamma \, r^2/2}  \right)$.

Figures 4-6 display the  $Y$-dependence of $\langle R_1(r, \theta; Y) \rangle_{\theta} $  behavior for the three ensembles,  each  consisting of ten complex matrices  of size $N=1024$ and considered for four $b$-values; (we note that the size ${\mathcal M}=10$ of the ensemble is sufficient for the analysis of $\langle R_1 \rangle$).  We numerically determine  $\langle R_1(r, \theta; Y) \rangle_{\theta}$  by counting the number of eigenvalues in an annular disk at a distance $r$, of width ${\rm d}r$ and centred at $r=0$ and then rescaling the number  by the area $2 \pi r {\rm d}r$.  The corresponding $(Y-Y_0)$ values in terms of $b$ for each case are obtained from eqs.(\ref{ybe}, \ref{ype}, \ref{yee}) using $b_0=1/N$, giving $Y-Y_0= -2 \,  {\rm ln} \;{ (1 + N/b) \over (1 + N^2)}$ for BE,  $= -{2\over N^2} \sum_ {r=1}^N  \, (N-r) \, \ln\left({1+r^2/ b^2 \over 1+N^2 r^2 } \right)$  for PE,  and,  $= {(N^2 b^2-1)\over N^2 b^2} \sum_ {r=1}^N  \, (N-r) \, r^2 = {(N^2 b^2-1)\, (N^2-1) \over 12 b^2} $  for EE.   
The specific $b$ values considered for each case and corresponding $(Y-Y_0)$ along with the fitted functions 
are given in table I.  (Consistent with our theory,  here we have  $(Y-Y_0) \ge 0$  for each case,  increasing from $0 \to \infty$ as $b$ varies from $1/N \to N$). 
We note, for the reference below,  that $(Y-Y_0)$ for EE case becomes very large for all $b$-values above $b=1/N$, thereby implying a rapid transition from the initial state to Ginibre limit.

Figures 4-6 also display a comparison with  eq.(\ref{rim2x}), eq.(\ref{rim2}) for small $b$-values  and eq.(\ref{rim1}) for $b=N$.  Although the $\sum_{\nu}$ includes infinite number of  terms,  we consider only first few terms  (with  $\nu \le 2$).  Further,  although the constants  $q_{1\nu}$ of $U_{1\nu}$ can in principle be determined from the initial condition,  the technical issues mentioned in {\it appendix} F leave us with no viable option but to determine them  by numerical fitting; the values for the fitting parameters in each case are given in table I.   As figure 4  for BE and figure 5 for PE indicate,  $\langle R_1(r, \theta; Y) \rangle_{r}$ varies slowly from its initial state $b=1/N$  and can be well fitted by eq.(\ref{rim2x})  for $b=1/N$ as well as for two intermediate values of $b$ with $\chi=0.001$;  (as mentioned near eq.(\ref{uka1}),  theoretically $\chi$ can be arbitrarily small). The details of the parameters for fitted functions for each case are given in table 1 (labelled as  $fit_{an}$). 

The figures 4 and 5  also indicate a good agreement with eq.(\ref{rim2}) for intermediate  $b$-values (again keeping only first few terms of $\sum_{\nu}$ referred by $fit_{bn}$);   this is expected due to slow change in $(Y-Y_0)$ for the chosen $b$-values for BE and PE.  Here  $\langle R_1(r, \theta; Y_0) \rangle_{r}$ required for  comparison is obtained by fitting the case $b=1/N$.  We note however that a good agreement with eq.(\ref{rim2})  for PE case,  occurs for a different $\chi (=0.7)$  value than the one used for fitting with eq.(\ref{rim2x}); a possible explanation of this deviation could be attributed to numerical stability.   (We also note that,  for the initial choice as $b=1/N$,  it is not appropriate to use a theoretical prediction  i.e a Poisson limit or the density of independent distributed eigenvalues on a complex plane for $\langle R_1(r,\theta,Y_0) \rangle_{\theta}$,   and it should  be determined numerically.  This is because BE, PE and EE, defined above eqs.(\ref{ybe}, \ref{ype}, \ref{yee}) respectively,  are expected to approach exact Poisson statistics only in $N \to \infty$ limit).  As expected on theoretical grounds, the behavior for the remaining case i.e $b=N$ for BE and PE  is  almost constant and is fitted by eq.(\ref{rim1}).   

Contrary to BE and PE,  the variation of   $\langle R_1(r, \theta; Y) \rangle_{r}$ with $b$ for EE case, displayed in figure 6,  is quite rapid.  Indeed it is well-fitted by eq.(\ref{rim1}) for all three cases except for the case $b=1/N$.   This is expected because  the $b$ dependence of $(Y-Y_0)$ for this case is almost negligible and  $(Y-Y_0)$ becomes very large  for all three values $b=10,  20,  N$.  This in turn also confirms  the sensitivity of the  spectral density to $(Y-Y_0)$ instead of $b$.

Proceeding similarly, $\langle R_1(r, \theta; Y) \rangle_{r}$ is numerically obtained, for a fixed $b$,  by counting the eigenvalues lying between sector $\theta$ and $\theta+{\rm d}{\theta}$ on the complex plane  and then rescaling the number  by the area $(1/2) r^2 {\rm d}{\theta}$ of the sector.  As predicted on theoretical grounds (discussed in section IV),  $\langle R_1(r, \theta; Y) \rangle_{r}$ is almost a constant for all considered $b$ values for each of the ensemble.   We emphasize that an agreement of our theoretical predictions  with fitted functional form for both $\langle R_1(r, \theta; Y) \rangle_{\theta }$ and $\langle R_1(r, \theta; Y) \rangle_{r}$ also lends credence to the separability ansatz made in section IV.

An important  prediction of our analysis,  mentioned in previous section and worth reemphasizing,  is as follows: based on the complexity parameter formulation,  $R_1$ can be described by the same mathematical form for a wide range of Gaussian ensembles  e.g. with varying degree of the sparsity if they  share the same $(Y-Y_0)$-value  and belong to same global constraints class although their local constraints may differ, resulting in different ensemble parameter values.  The analogy however is not limited to a single static point on the evolutionary path; if a random  perturbation subjects the ensemble to evolve, it  continues lying along the same evolutionary path  in the ensemble  space (constrained by  fixed values of $t_2, \ldots, t_M$ based on  global  constraints).   As each case displayed in figures 4-6 is well-described by either eqs.(\ref{rim2x},  \ref{rim1}) obtained from complexity parameter formulation of the spectral density,  the above prediction is indeed well supported by our numerical results.


\section{Conclusion}

In the end, we summarize our main ideas and results discussed above along with some open questions.

Based on the representation by a multiparametric Gaussian ensemble of complex matrices,  we find that the ensemble averaged spectral density of a non-Hermitian complex system is sensitive to system-specific details only through the complexity parameter, a single functional of the system conditions.  A variation of system conditions may subject the spectral density to evolve in spectral as well as ensemble space, with global constraints e.g. conservation laws and symmetries act as the constants of dynamics.    Our search for a  path in the ensemble space, the evolution along which  mimics that on the spectral plane,  leads  to  a single parametric "universal" path fixed by a set of global constraints.  The  relevance of such a path can be explained as follows:  when system parameters vary,  the ensembles,   representing different systems subjected to same set of fixed global constraints,  are confined to evolve along this path.   This also describes a universal dynamics, in terms of the complexity parameter $Y$, for the eigenvalues of complex matrices (with Gaussian randomness)  in an arbitrary initial state and subjected to mutiparametric perturbations equilibrium; the   dynamics approaches the steady state of Ginibre ensemble for large $Y$-limits.  More explicitly, 
as different  complex matrix ensembles occurring at same point of this path (i.e those with same value of $Y$ and similar initial statistics) have same statistical spectral properties,  this indicates the hidden universality even in non-equilibrium regime  i.e beyond Ginibre ensemble.

An important result of our analysis is the solution of  the evolution equation for the average spectral density for arbitrary values of the complexity parameter.    Although  technical complexity of the differential equation made it necessary to assume the weak second order spectral correlations on the complex plane,  the conjecture is based  on  the very definition of the local spectral correlations,   expected to be weaker beyond a single local mean level spacing and is therefore valid for calculation of the average spectral density.  Further, due to arbitrary choice of the ensemble parameters in our analysis,  the solution is applicable to a wide range of ensembles e.g. various type of sparse matrix structures that describes the statistical behavior of non-Hermitian operators of complex systems e.g. many body or disordered  systems.    The above claim is indeed confirmed by the numerical analysis of three ensembles with different degree of the sparsity of the off-diagonals; we find that their numerically obtained average spectral densities agree well with our theoretical  formulation in terms of the complexity parameter and based on separability conjecture.

We note that our present numerical analysis is based  on the choice of  initial condition corresponding to Poisson spectral statistics on a complex plane and is chosen due to  intense current interest in the Poisson to Ginibre transition in spectral statistics,  with intermediate state described by  various sparse random matrix ensembles representing many real non-Hermitian systems.   But as our evolution equations for the {\it jpdf} of eigenvalues,  and thereby spectral density,  are valid for any arbitrary initial condition,  it is desirable to pursue a numerical verification for other initial conditions too.   For example,  the choice of an Ginibre ensemble of real asymmetric matrices as an initial condition would lead to a crossover/ transition between two different universality classes  of Ginibre ensembles on complex spectral plane.  Similarly the choice of any other universality classes mentioned in \cite{hkku}  as an initial condition would give rise  to  a different transition from the chosen initial statistics to that of Ginibre ensemble of complex matrices.  Such studies are indeed expected to give new insights in the ergodic/ non-ergodic aspects of the spectral statistics. 

Our study also gives rise to many other  queries e.g. can the complexity parameter formulation and the universality in non-equilibrium regime be extended to non-Gaussian ensembles as well as to structured ensembles with correlated matrix elements? The information is  relevant in order to model more generalized class of  non-Hermitian systems where existence of additional matrix constraints can lead to many types of matrix elements correlations. Another very important question for application to many body non-Hermitian systems is how the behavior of correlations in the edge differ from the bulk? The information is needed to understand e.g. the non-Hermitian skin effect and the absence of conventional bulk-boundary correspondence in topological and localization transitions in non-Hermitian systems \cite{am} and also for applications to biological neural networks.  Another  natural future extension of our analysis is to seek the explicit solution for the complexity parameter based  evolution equation  for the higher order  spectral correlations on the complex plane.  A similar formulation for the eigenvector correlations  if possible is also desirable.  We intend to answer some of the above queries in near future.

\acknowledgments

One of the authors (P.S.)  is grateful to SERB, DST, India for the financial support provided for the  research under Matrics grant scheme.


\newpage

\appendix

\section{Complexity parameter formulation of $\rho$}

As discussed in section II. A,   we seek a transformation from the set of $M=2 N^2$ parameters $\{y_{kl;s}, x_{kl;s} \}$  (with $x_{kl;s} = x_{lk;s}$) to another set $\{t_1,\ldots, t_M \}$,  such that multiparametric evolution governed by the operator $T$, defined in eq.(\ref{trho}), can be reduced to a single parameter evolution, say $t_1$ while rest of them i.e $t_2, \ldots, t_M$ remain constant.

We redefine, for technical ease,    the ensemble density $\tilde\rho$ as $\rho_1= C_2 \rho = {C_2 \over C} \tilde\rho$, with $C$ as the normalization constant of $\tilde\rho$ (such that $\int  \tilde\rho {\rm d}H =1$).   Here  $C_2$ is an unknown function of the parameters $y_{kl;s}$ and $x_{kl,s}$  such that $\rho$ satisfies the condition 
\begin{eqnarray}
T \, \rho +C_1 \rho = T \, \rho_1 = {\partial \rho_1\over\partial t_1}
\label{diff1a}
\end{eqnarray}

The above in turn gives the condition to determine $C_2$:   
\begin{eqnarray}
T \, C_2 - C_1 C_2  = 0. 
\label{diff3a}
\end{eqnarray}

To fulfill the second part of the condition in eq.(\ref{diff1a}) i.e $ T \, \rho_1 = {\partial \rho_1\over\partial t_1}$,   $t_k$  must satisfy following condition 
\begin{eqnarray}
 \sum_{s=1}^{\beta}  \sum_{k,l}
 \left[ A_{kl;s} {\partial t_1 \over\partial y_{lk;s}}
 +  B_{kl;s}  {\partial t_1 \over\partial x_{kl;s}}\right] = 1.
\end{eqnarray}
and 
\begin{eqnarray}
 \sum_{s=1}^{\beta}  \sum_{k,l}
 \left[ A_{kl;s} {\partial t_k \over\partial y_{lk;s}}
 +  B_{kl;s}  {\partial t_k \over\partial x_{kl;s}}\right] = 0,   \hspace{0.2in}   k >1
\end{eqnarray}
with  $A_{kl;s}, B_{kl;s}$ defined in eq.(\ref{dab}).

The above set of equations can be solved by standard method of characteristics; the latter leads to a set of $M$ relations

\begin{eqnarray}
 {{\rm d}y_{11;1} \over A_{11;1}}
= {{\rm d}x_{11;1} \over B_{11;1}}
= \ldots 
= {{\rm d}y_{kl;s} \over A_{kl;s}}
= {{\rm d}x_{kl;s} \over B_{kl;s}}
={{\rm d}t_1 \over 1}, 
\label{t1}
\end{eqnarray}
and 
\begin{eqnarray}
 {{\rm d}y_{11;1} \over A_{11;1}}
= {{\rm d}x_{11;1} \over B_{11;1}}
= \ldots 
= {{\rm d}y_{kl;s} \over A_{kl;s}}
= {{\rm d}x_{kl;s} \over B_{kl;s}}
={{\rm d}t_k \over 0}   \qquad  k > 1.
\label{tk}
\end{eqnarray}

 Substitution of $A_{kl;s}$ and $B_{kl;s}$ from eq.(\ref{dab}) and rearranging eq.(\ref{t1}), we have, with $x_{kl;s}=x_{lk;s}$,
\begin{eqnarray}
{{\rm d}h_{lk;s} \over h_{lk;s} - 2 x^2_{kl;s}} =  {{\rm d}h_{kl;s} \over h_{kl;s} - 2 x^2_{kl;s}}= {{\rm d}x_{kl;s} \over x_{kl;s}} 
\label{t2}\end{eqnarray}
where $h_{kl;s} = y_{kl;s} (\gamma - 2 y_{kl;s})$. 
Solving the above set of equality relations in eq.(\ref{t2}), we have 
\begin{eqnarray}
x_{kl;s} &=& \tilde c_{kl;s} (h_{kl;s} - h_{lk;s}) \\
 y_{lk;s} &=&  c_{kl;s} \; y_{kl;s}\label{t3}
 \end{eqnarray}
Substitution of $x_{kl;s}$ in eq.(\ref{t1}), we now have 
\begin{eqnarray}
 {{\rm d}y_{11;1} \over F_{11;1}}= \ldots = {{\rm d}y_{kl;s} \over F_{kl;s}}={{\rm d}t_1 \over 1}, \label{t8}\end{eqnarray}
with $F_{kl;s} = y_{kl;s} \, w$ where $w= 2 \gamma - 4 y_{kl;s} - \tilde c_{kl;s}^2 (1-c_{kl;s})^2 y_{kl;s} (\gamma - 2 (1+c_{kl;s}) y_{kl;s})^2$. The above on solving now gives 
\begin{eqnarray}
t_1 
&=& \pm \, {1\over Q} \sum_{k,l;s}  a_{kl;s} \; \int {{\rm d}y_{kl;s} \over y_{kl;s} \, w} + C_0 \\
\label{yt1a} 
\end{eqnarray}
where $Q=\sum_{kl;s} a_{kl;s}$ with $a_{kl;s}$ and $C_0$ given by the initial conditions.  Under condition that all $M/2$ ensemble parameters are undergoing variation during evolution,one can set all  $a_{kl;s}$ equal e.g $a_{kl;s}=1$ or $a_{kl;s}=-1$ thereby giving $Q=\pm N^2$.



Similarly solving set of $M-1$ eq.(\ref{tk}) lead to $M-1$ constants of motion $t_k$.  As a similar approach for Hermitian operators has been discussed in a series of studies \cite{psco, psall}, and in \cite{psnh} for non-Hermitian operators, the details of the derivation are not included in this work.

\section{Derivation of eq.(\ref{pcmp})}

Let  $\tilde P(z,y,x) $ be the probability of finding eigenvalues
$\lambda_i $  of $H$ between $z_i$ and $z_i+{\rm d}z_i$ for  
 given sets $y$ and $x$,
\begin{eqnarray}
\tilde P(z,y,x) =  \int f(z,z^*) \; \tilde \rho (H,y,x) \, {\rm d}H 
\label{pe0}
\end{eqnarray}
with $z \equiv \{z_i\}$ and $f(z,z^*) = \prod_{i=1}^{N}\delta(z_{i}-\lambda_{i})
\delta(z_i^*-\lambda_i^*)$ or equivalently .$f(z,z^*) = \prod_{i=1}^{N}  \prod_{b=1}^{2}\delta(z_{ib}-\lambda_{ib})$. with 
$z_i = z_{i1}+i z_{i2}$.  

For technical ease,  we  define $P_1(z, y,x)  ={C_2\over C}  \tilde P(z,y,x)$.  Using $\rho_1 ={C_2\over C}  \tilde \rho$, 
this leads to
\begin{eqnarray}
P_1(z,y,x) =  \int f(z,z^*) \; \rho_1(H,y,x) \, {\rm d}H 
\label{pe1}
\end{eqnarray}
 
Differentiating eq.(\ref{pe1}) both sides with respect to $Y \equiv t_1$ 
leads to 
\begin{eqnarray}
{\partial  P_1\over \partial Y} =  \int f(z,z^*) \, {\partial  \rho_1\over \partial Y}  {\rm d}H 
\label{pe2}
\end{eqnarray}
A substitution of eq.(\ref{rhot1}) in the above integral now gives
\begin{eqnarray}
{\partial  P_1 \over \partial Y} = \gamma \;  I_1  + I_2
\label{pe3}
\end{eqnarray}
where 
\begin{eqnarray}
 I_1 &=& \sum_{k,l;s} \int f \, {\partial  ( H_{kl;s}  \rho_1) \over \partial H_{kl;s}} \; {\rm d}H  \\
I_2  &=& \sum_{k,l;s} \int f \, {\partial^2   \rho_1\over \partial H_{kl;s} \partial  H_{kl;s}} \; {\rm d}H   
\label{pe4}
\end{eqnarray}
Using repeated partial integration  and the boundary condition that $\rho$ vanishes as $H_{kl;s} \to  \pm \infty$
$I_1$ and $I_2$ can  be rewritten as 
\begin{eqnarray}
 I_1  &=& \sum_{r,n} {\partial \over \partial z_{nr}}
\int f   \;  \left(\sum_{k,l;s} {\partial \lambda_{nr} \over \partial H_{kl;s}}
H_{kl;s} \; \rho_1 \right)  \;{\rm d}H
\label{i1}
\end{eqnarray}
and 
\begin{eqnarray}
 I_2  &=& \sum_{r,n} \sum_{b,m}{\partial^2 \over \partial z_{nr} \partial z_{mb}}
\int f   \;  \left(\sum_{k,l;s} {\partial \lambda_{nr} \over \partial H_{kl;s}} {\partial \lambda_{mb} \over \partial  H_{kl;s}} \right)
 \; \rho_1  \;{\rm d}H   + \sum_{r,b, m, n} {\partial \over \partial z_{nr}}
\int f   \;  \left(\sum_{k,l;s} {\partial^2 \lambda_{nr} \over \partial  H_{kl;s} \partial H_{kl;s}} \right)  \; \rho {\rm d}H
\nonumber  \\
\label{i2}
\end{eqnarray}

To proceed further, we need the information about the response of the eigenvalues as well as the eigenvectors to a small change in the matrix element $H_{kl}$.  Using the eigenvalue equation $\Lambda = U H V$ with $U$ and $V$ as unitary matrices,  the required  relations can be given as follows 

\begin{eqnarray}
{\partial \lambda_n \over \partial H_{kl;s}} =
   i^{s-1} U_{nk} V_{ln};  \hspace{0.1in}
\sum_{k,l;s} {\partial \lambda_n \over\partial H_{kl;s}} H_{kl;s}
 = \lambda_n, \nonumber \\ 
  \sum_{k,l;s} {\partial \lambda_n \over\partial H_{kl;s}}
  {\partial \lambda_m^* \over\partial H_{kl;s}}  = \beta \delta_{mn},
\nonumber \\
 \sum_{k,l;s} 
 {\partial^2 \lambda_n \over\partial H_{kl;s}^2}
= \sum_{m} {\beta \over \lambda_n^* - \lambda_m^*} \nonumber \\
{\partial U_{ni} \over\partial H_{kl;s}} =
(-i)^{s-1} \,  \sum_{m\not=n}
{U_{mi} \, U_{mk}^* \, V_{ln}^* \over {\lambda_n^* -\lambda_m^*}},\nonumber \\
{\partial V_{jn} \over\partial H_{kl;s}} = (i)^{s-1}
 \sum_{m\not=n}{V_{jm} V_{ln} U_{mk}\over {\lambda_n -\lambda_m}}
\label{evf}
\end{eqnarray}


Using  the above relations in eqs.(\ref{i1}, \ref{i2}),  $I_1$  and $I_2$ can be rewritten as 
$I_1= \sum_{n,r} {\partial \over \partial z_{nr}} \left(z_{nr} P_1\right)$
and 
$I_2=  {\partial^2 P_1\over \partial z_{nr}^2}
- 2 {\partial \over \partial z_{nr}}
\left( {\partial {\rm ln} |\Delta (z)| \over \partial z_{nr}} P_1\right) $ with
  $\Delta_N(z)=\prod_{j<k}^N(z_j-z_k)$.   
A substitution of these equalities in eq.(\ref{pe3})  now
 gives  the diffusion equation (\ref{pcmp}) for $P_1$.

\section{Derivation of eq.(\ref{r0z})}

An integration of eq.(\ref{pcmp}) over the variables $z_{2}, \ldots, z_N$ (as well as their conjugates), multiplying both sides by $N$ and subsequently using eq.(\ref{rn0})  leads to 
\begin{eqnarray}
 {\partial R_1\over \partial Y} &=& N \int  \,  {\partial P_1\over \partial Y} \,   {\rm  d}^2 z_{2}, \ldots, {\rm  d}^2 z_N  \nonumber \\
&=& \sum_{r=1}^2 \sum_{n=1}^N  \left( G_{nr;1} +G_{nr;2} +G_{nr;3} \right)
\label{rp0}
\end{eqnarray}
where 
\begin{eqnarray}
G_{jr;1} &=& N \, \int  \,  {\partial^2 P_1\over \partial z_{jr}^2} \,   {\rm  d}^2 z_{2}, \ldots, {\rm  d}^2 z_N  ,  \nonumber \\
&=& {\partial^2 R_1\over \partial z_{1r}^2}  \hspace{2.2in}  j =1, \label{rpp1} \\
&=&  0 \hspace{2.5in} j > 1,
\label{rp1}
\end{eqnarray}
with ${\rm  d}^2 z \equiv {\rm  d} z \, {\rm  d}^* z$.  The second relation in the above equation follows by noting that,  for $j=1$,  ${\partial^2 \over \partial z_{1r}^2}$ can be taken out of the integral; subsequent use of eq.(\ref{rn0}) then gives eq.(\ref{rpp1}).  For $j \not=1$, the repeated partial integration over $z_j$ along with the boundary condition that $P(z_j)\to 0$ as $z_j \to \pm \infty$ gives eq.(\ref{rp1}).

Similarly
\begin{eqnarray}
G_{jr;2} &=& - 2 \; N \, \int {\partial \over \partial z_{jr}} \left( \ln|\Delta_N(z)| \, P_1\right) \; {\rm  d}^2 z_{2}, \ldots, {\rm  d}^2 z_N, \nonumber \\
&=& - {2 \, N\over N-1} \, {\partial \over \partial z_{j1}} \int {z_{1r}-z_{2 r}\over |z_1-z_2|^2} \, R_{2}(z)\, {\rm d^2}z_{2} \qquad   j=1,  \nonumber \\
&=&  0  \hspace{2.85in}  j > 1
\label{rp2}
\end{eqnarray} 
with $R_{2}(z) \equiv R_{2}(z_1,  z_2)$.
 and 
 \begin{eqnarray}
G_{jr;3} &=& N \, \int  \,   {\partial \over \partial z_{jr}} \left(z_{jr} \, P_1\right)  \,{\rm  d}^2 z_{2}, \ldots, {\rm  d}^2 z_N \nonumber \\
 &=& {\partial \over \partial z_{1r}} \left(z_{1r} \, R_1\right), 
\qquad j=1, \nonumber \\
&=& 0  \hspace{1.2in} j > 1
\label{rp3}
\end{eqnarray} 
Substitution of eqs.(\ref{rp1}, \ref{rp2}, \ref{rp3}) in eq.(\ref{rp0}) then leads to eq.(\ref{r0z}).

\section{Derivation of eq.(\ref{r1rt})}
With $z=x+i y$, $z'=x'+i y'$, $R(z)=R(x, y)$, eq.(\ref{r0z}) can be rewritten as 
\begin{eqnarray}
{\partial R_1\over \partial Y} = I_0  + \gamma \, I_1 +  I_2
\label{r1rta}
\end{eqnarray}
 where $ I_0 =  {\partial^2 R_1 \over \partial r^2} + {1\over r^2} \; {\partial^2  R_1 \over \partial \theta^2}  + {2\over r} \, {\partial R_1 \over \partial r} $
and
\begin{eqnarray}
I_1 = {\partial  (x R_1)\over \partial x} + {\partial  (y R_1)\over \partial y}   \label{i1a}
\\
I_2 = {\partial  (G_x )\over \partial x} + {\partial  (G_y )\over \partial y}   \label{i2a}
\end{eqnarray}
with 
\begin{eqnarray}
G_x &=& - 2 \; {\bf P}\int {d}x' \, {d}y' \, R_2 (z, z')  \, {x- x'\over |z- z'|^2} \label{gx} \\
G_y &=& - 2 \; {\bf P}\int {d}x' \, {d}y' \, R_2 (z, z')  \, {y- y'\over |z- z'|^2} \label{gy}
\end{eqnarray}

Using standard transformation rules from cartesian to polar coordinates i.e $x=r \, \cos \theta$, $y=r \, \sin \theta$, we have ${\partial \over \partial x} = \cos \theta \, {\partial \over \partial r} - {\sin \theta\over r} \, {\partial \over \partial \theta}$, 
${\partial \over \partial y} = \sin \theta \, {\partial \over \partial r} + {\cos \theta\over r} \, {\partial \over \partial \theta}$, we then have
\begin{eqnarray}
I_1  &=&  r \, {\partial R_1 \over \partial r} + 2 \, R_1  
\label{i1ia}\\
I_2 &=&   {\partial I_c \over \partial r} + {1\over r} \,  {\partial I_d \over \partial \theta} + {I_c \over r} 
\label{i2ia}
\end{eqnarray}
where $r \, I_c \equiv G_x \cos \theta + G_y \sin \theta $ and $I_d \equiv G_x \sin \theta - G_y \cos \theta $. Writing $x, y$ in term of $r, \theta$ and using  ${d}x' \, {d}y' = r' \, {d}r' \, {d}\theta'$  and $|z-z'|^2 = r^2+ r'^2-2 \, r \, r' \, \cos(\theta-\theta')$
 in eqs.(\ref{gx}, \ref{gy}), $I_c$ and $I_d$ can now be rewritten as given in eq.(\ref{ac}) and eq.(\ref{ad}). Substitution of eq.(\ref{i1ia}) and eq.(\ref{i2ia}) in eq.(\ref{r1rta}) then gives eq.(\ref{r1rt}).

\section{Determination of $a_{\mu \nu}$ from a given  initial spectral density}

As our solution of the dynamical equation for $R_1(r, \theta; Y)$ discussed in section IV  is $r$-range specific and therefore gives different 
conditions to determine unknown coefficients  for small,  large and finite $r$ regimes but with decaying angular correlations.   As discussed below,  here we consider two cases of initial conditions as examples.

(i) {\it Gaussian decay of initial density with circular symmetry:}

As an example, we consider  the initial ensemble with 
\begin{eqnarray}
R_1(r, \theta; Y_0)  = {1\over \sqrt{\pi}} \, {\rm e}^{- q r^2}
\label{ex1}
\end{eqnarray}

{ {\it Small} $ r $ :} Substitution of the above in eq.(\ref{dsol}) and using $Y=Y_0$ leads to the condition, valid only for $r \le {1\over \sqrt{N}}$,

\begin{eqnarray}
{1\over \sqrt{\pi}} \, {\rm e}^{-q \, r^2} = \int_0^{\infty} {\rm d}E \; a(E)  \,   r^{-1/2} \, J_{1/2} \left( r\sqrt{E} \right).
\end{eqnarray}

  Using the definition $J_{\mu}(x)= \left({x \over 2}\right)^{\mu} \; \sum_{k=0}^{\infty} (-1)^k {1\over k! \Gamma(\mu+k+1)} \; \left({x\over 2}\right)^{2k} \approx \sqrt{x\over 2 \pi} $ for $x \to 0$,  $a(E)$ satisfies the relation  $\int_0^{\infty}  {\rm d}E \;  a(E) \; E^{1/4} = \sqrt{2}$ and  can now be obtained by an inverse Mellin transform.

{ {\it Large} $ r$ :} To determine unknown coefficients in our large $r$-solution given by eq.(\ref{ri}),  we  now have the condition 
\begin{eqnarray}
{1\over \sqrt{\pi}} \, {\rm e}^{-q r^2} = {1\over \langle r \rangle} \,  \,  \sum_{\mu,  \nu}  \; b_{ \mu }  \,  U_{\mu \nu}  \; {\rm e}^{ - {(\mu^2-1) (2 N-1)^2 \, I_0 \over 4 \langle r \rangle} }
\end{eqnarray}
 with $U_{\mu \nu}$ given by eq.(\ref{umnn}) with $N_0=2 N$ and $I_0$ defined in eq.(\ref{ti0}).  
Clearly the condition can be fulfilled only if $\mu=1$ and therefore we must have $a_{\mu \nu}=b_{\mu} \, d_{\mu \nu} =0$ for $\mu >1$. This in turn reduces the above as
\begin{eqnarray}
{1\over \sqrt{\pi}} \, {\rm e}^{-q r^2}  &=& {1\over \langle r \rangle} \, \left({ \gamma \, r^2\over 2}\right)^{\frac{(2 N-1)}{2}} \, \sum_{\nu}    \,  a_{1\nu} \; F_1\left((2\nu+1) N, \ N, -\frac{\gamma r^2}{2}\right)   \nonumber \\
&\sim& {1\over \langle r \rangle} \,  \sum_{\nu}    \,  a_{1\nu} \; {\rm e}^{-(2\nu+1) \frac{\gamma r^2}{2}}. 
\label{exb1}
\end{eqnarray}
A comparison of the two sides of the above equation now gives $a_{10} ={ \langle r \rangle \over \sqrt{\pi}}$ and $a_{1 \nu} =0$ for $\nu >0$ and $\gamma=2 \, q$.

 {\it Finite} $ r$ : Using eq.(\ref{rij}) for $Y=Y_0$ along with eq.(\ref{rina}) ),  eq.(\ref{rij}) gives   for case C.(i),
\begin{eqnarray}
{1\over \sqrt{\pi}} \, {\rm e}^{-q r^2}
&= &   \sum_{\mu, \nu} \, b_{\mu}  \;  U_{\mu \nu} \; \cos\left(\frac{ \alpha_1}{2} \sqrt{\mu^2-1} \, \theta \right)
 \label{rjae}
\end{eqnarray}
with $U_{\mu \nu}$ given by eq.(\ref{uia1}), eq.(\ref{uja1}) and eq.(\ref{uka1}) based on $f(r)$.  With its right side  $\theta$-dependent and the left side  independent,  the  condition eq.(\ref{rjae}) can  be fulfilled only for $\mu=1$.  The latter leads to the condition  ${1\over \sqrt{\pi}} \, {\rm e}^{-q r^2} = b_1 \,  \sum_{\nu}   U_{1 \nu} $.

The standard route to determine constants from equations such as above is based on using orthogonality of functions appearing in the series. However as the same are not known in case of confluent Hypergeometric functions,  one option left to us is to expand both sides in power series and compare the terms on both sides.

(ii) {\it Case for initial density used in numerical analysis: }

The initial condition used for our numerical analysis correponds to $b=1/N$ for each of the three ensembles. 
As displayed in figures 4-6 for cases $b=1/N$ for BE, PE and EE,   the spectral density in this case is as follows:
\begin{eqnarray}
R_1(r, \theta; Y_0) = A \, r^{-1/2} \; J_{1/2}(B r) \, {\rm e}^{-C r^2}.   \label{rina}
\end{eqnarray}

As in the previous case,  here again we consider three different regimes to determine the coefficients.

{ {\it Small} $ r$ :}  Using eq.(\ref{dsol}), for $Y=Y_0$ along with eq.(\ref{rina}) (with  ${\rm e}^{-\phi r^2}. \approx 1$), we have,
\begin{eqnarray}
A \, r^{-1/2} \; J_{1/2}(B r)  &=&   r^{-1/2} \, \int_0^{\infty}  {\rm d}E \; a(E) \, J_ {1/2}\left( r \sqrt{E} \right)   
\label{dsola1}
\end{eqnarray}
Solving the above now gives $a(E)=\delta(E-\beta^2)$. 

{{\it Large}  $ r $ :}  Using eq.(\ref{ri}), for $Y=Y_0$ along with eq.(\ref{rina}) (with  $J_{1/2}(B r) = {\sin(B r)\over \sqrt{\pi B r}}$), we have  for large $r$,
\begin{eqnarray}
 {A \over \sqrt{\pi B r^2}} \; \sin(B r) \; {\rm e}^{- C r^2}.   
& \approx & {1\over \langle r \rangle} \, \left({ \gamma \, r^2\over 2}\right)^{\frac{(2 N-1)}{4}} \, \sum_{\mu, \nu} \,  a_{\mu \nu} \, U_{\mu \nu}(r)  \; \, {\rm e}^{ - {(\mu^2-1) (2 N-1)^2 \, I_0(\theta) \over 4 \langle r \rangle} }  
\label{ria} 
\end{eqnarray}
with $U_{\mu \nu}$ given by eq.(\ref{umnn}).  
As the left side of eq.(\ref{ria}) does not depend on $\theta$,   this implies $a_{\mu \nu} =0$ for $\mu >1$.  The condition in eq.(\ref{ria}) now reduces as
\begin{eqnarray}
 {A \over \sqrt{\pi B r^2}} \; \sin(B r) \; {\rm e}^{-C r^2}.   
& \approx & {1\over \langle r \rangle} \, \left({ \gamma \, r^2\over 2}\right)^{\frac{(2 N-1)}{4}} \, \sum_{\nu} \,  a_{1 \nu} \, U_{1 \nu}(r)    
\label{ria1} 
\end{eqnarray}

{\it Finite  $  r$ :}  Using eq.(\ref{rij}) for $Y=Y_0$ along with eq.(\ref{rina}) and  $J_{1/2}(B r) = {\sin(B r)\over \sqrt{\pi B r}}$),  eq.(\ref{rij}) now gives   for case C.(i),
\begin{eqnarray}
 {A \over \sqrt{\pi B r^2}} \; \sin(B r) \; {\rm e}^{-C r^2}
&= &   \sum_{\mu, \nu} \, b_{\mu}  \;  U_{\mu \nu} \; \cos\left({1\over 2} \sqrt{\mu^2-1} \, \alpha_1 \, \theta \right)
 \label{rja}
\end{eqnarray}
with $U_{\mu \nu}$ given by eq.(\ref{uia1}), eq.(\ref{uja1}) and eq.(\ref{uka1}) based on $f(r)$.  Here again $\theta$-independence the left side of eq.(\ref{rja})  requires  $b_{\mu} =0$ for $\mu >1$.   The condition in eq.(\ref{rja}) now reduces to  $ {A \over \sqrt{\pi B r^2}} \; \sin(B r) \; {\rm e}^{-C r^2} = b_1 \,  \sum_{\nu}   U_{1 \nu} $.  The constants in $U_{1\nu}$ can again be determined,  in principle,  by inverting the condition with help of orthogonality relations of the mathematical functions appearing in $U_{1\nu}$ or by expanding both sides as a power series and comparing the terms with same powers.

\section{Calculation of $\langle R_1(r,  \theta, Y) \rangle_{\theta}$ and $\langle R_1(r,  \theta, Y) \rangle_{r}$}

Using the functions  $\langle R_1 \rangle_{\theta}=\int_0^{2\pi} R_1 \, {\rm d}\theta$ and  $\langle R_1\rangle_{r}=\int_0^{\infty} R_1 \, r \, {\rm d}r$, the radial and angular dependence of $R_1$  can separately be analyzed.

{\bf Radial Dependence:}  Using the definition  eq.(\ref{rij})   we have

\begin{eqnarray}
\langle R_1(r, \theta,Y) \rangle_{\theta}  &=&   \sum_{\nu}    \, U_{1\nu}  \, {\rm e}^{- \nu \, \phi }   
 \label{rima}
\end{eqnarray}
with $\phi= 4\gamma  \alpha (Y-Y_0)$ for the case $f(r)= \frac{\alpha}{r}$ and $f= \alpha \, r$ and $\phi= 2 \gamma \chi (Y-Y_0)$ for the case $f(r)$ decaying exponentially or faster.   

For small $\phi $,  we can approxmate ${\rm e}^{- \nu \phi }  \approx 1 - \nu \phi + O(\phi^2)$.  This implies, for small $Y-Y_0$,

\begin{eqnarray}
& &\langle R_1(r,\theta,Y) \rangle_{\theta}  
\approx  \langle R_1(r,\theta,Y_0) \rangle_{\theta} - \phi \;  \sum_{ \nu=1}^{\infty}  \nu \; U_{1\nu} \\
&\approx & \langle R_1(r,\theta,Y_0) \rangle_{\theta} \;  \left(1- \phi \right)  \; + \phi  \, U_{10} -S_2 
\label{rio3a} 
\end{eqnarray}
where $S_2 \equiv \phi \;  \sum_{ \nu=2}^{\infty}  (\nu-1) \; U_{1 \nu} $.

For  $Y > Y_0$,    the terms with increasingly lower values of $\nu$ begin to dominate the above sum, with  only $\nu=0$ term contributing in $Y-Y_0 \to \infty$.   For large $Y-Y_0$, the above series can then  be approximated by retaining only some lower values of  $\nu$ i.e

\begin{eqnarray}
\langle R_1(r, \theta,Y) \rangle_{\theta}  & \approx &     U_{10} +   \, U_{11}  \, {\rm e}^{- \phi }   
 \label{rimaa1}
\end{eqnarray}

{\it Determination of coefficients:}  Eq.(\ref{rima}) gives   $\langle R_1(r, \theta,Y_0) \rangle_{\theta}  =  \sum_{\nu}    \, U_{1\nu}  $; the unknown constants  $p_{1\nu}$ and $q_{1\nu}$ of $U_{1\nu}$ can then be determined from the initial condition.
For example,  for initial density given by  eq.(\ref{rina}), we have (using eq.(\ref{uka1}))
\begin{eqnarray}
{A \over \sqrt{r}} \; J_{1\over 2}(B r) \, {\rm e}^{-C r^2} =  \sum_{ \nu=0}^{\infty}  \; \left[ p_{1 \nu} \, \left({ \gamma \, r^2 \over 2}\right)^{-\frac{1}{2}} \; F_1\left(\frac{1}{2}+\nu \chi,  \frac{1}{2}, \,  -\frac{\gamma r^2}{2}\right)  + q_{1 \nu} \, \, F_1\left( \nu \chi +1,  \frac{3}{2},-\frac{\gamma r^2}{2}\right)  \right] \nonumber \\
 \label{rinaa3a}
\end{eqnarray}

Using $J_{1/2}(B r) = {\sin(B r)\over \sqrt{\pi B r}}$ and expanding both sides in a power series near $r=0$ and  comparing various powers of $r$, we have  $p_{1 \nu}=0$ and
\begin{eqnarray}
z_n  =   \sum_{ \nu=0}^{\infty}  \;  q_{1\nu}  \, (\nu \chi +1)_n
 \label{rina4a}
\end{eqnarray}
where $z_{n} =\frac{A}{\sqrt{\pi}} \; \left(\frac{2}{\gamma}\right)^n  \left(\frac{3}{2}\right)_n \, \sum_{k=0}^{n}  {n!  B^{2k+1/2} C^{n-k} \over (n-k)!  (2k+1)!}$.   Here $(a)_n$ is a Pochhammer symbol i.e $(a)_n = a(a+1)...(a+n)$.

Eq.(\ref{rina4a}) can further be written in form of  a matrix equation $Z =  W Q$   with $Z$ and  $Q$  as the column vectors,  $Z \equiv [z_n]$, $Q_{\nu} \equiv [q_{1 \nu}]$ (both $n, \nu=1 \to \infty$)
and $W$  as an infinite dimensional  matrix with entries $W_{\nu n} =(\nu \chi +1)_n$.  Inverting the above  gives  $Q = W^{-1} \; Z$ and thereby $q_{1 \nu}$ for each $\nu$.   In  practice however inversion is not easy due to infinite dimansionality of the matrices involved, leaving the determination of the constants $q_{1\nu}$ only by numerical fitting.

\vspace{0.5in}

{\bf Angular Dependence:}  From eq.(\ref{rij}),  we also have
\begin{eqnarray}
\langle R_1(r, \theta,Y) \rangle_{r}  \approx   \sum_{\mu=1}^{\infty} \cos\left((1/2) \sqrt{\mu^2-1} \, \alpha_1 \, \theta \right) \; X(\mu, Y) 
\label{rim3a} 
\end{eqnarray}
where $X(\mu; Y)$ is independent of $\theta$: $X(\mu, Y) = \sum_{ \nu} \,   b_{\mu } \, H_{\mu \nu}   \, {\rm e}^{- 2 \gamma \, \nu  N  (Y-Y_0) } $  with $ H_{\mu\nu}=\int_0^{\infty} U_{\mu \nu}  \; r \;   {\rm d}r $ and $U_{\mu \nu}$ dependent on $f$ (eq.(\ref{j000})).   As clear from the above,   the dominant contribution to eq.(\ref{rim3a}) comes from $\mu=1$ term, thereby suggesting $\langle R_1(r, \theta,Y) \rangle_{r}  \approx  X(1, Y)$;  this is consistent with our numerical results,  displayed in figures 4-6,  indicating $\langle R_1(r, \theta; Y) \rangle_{r}$   almost constant  in $\theta$.

\begin{figure}[ht!]
\centering

\vspace{-1in}

\includegraphics[width=16 cm,height=22 cm]{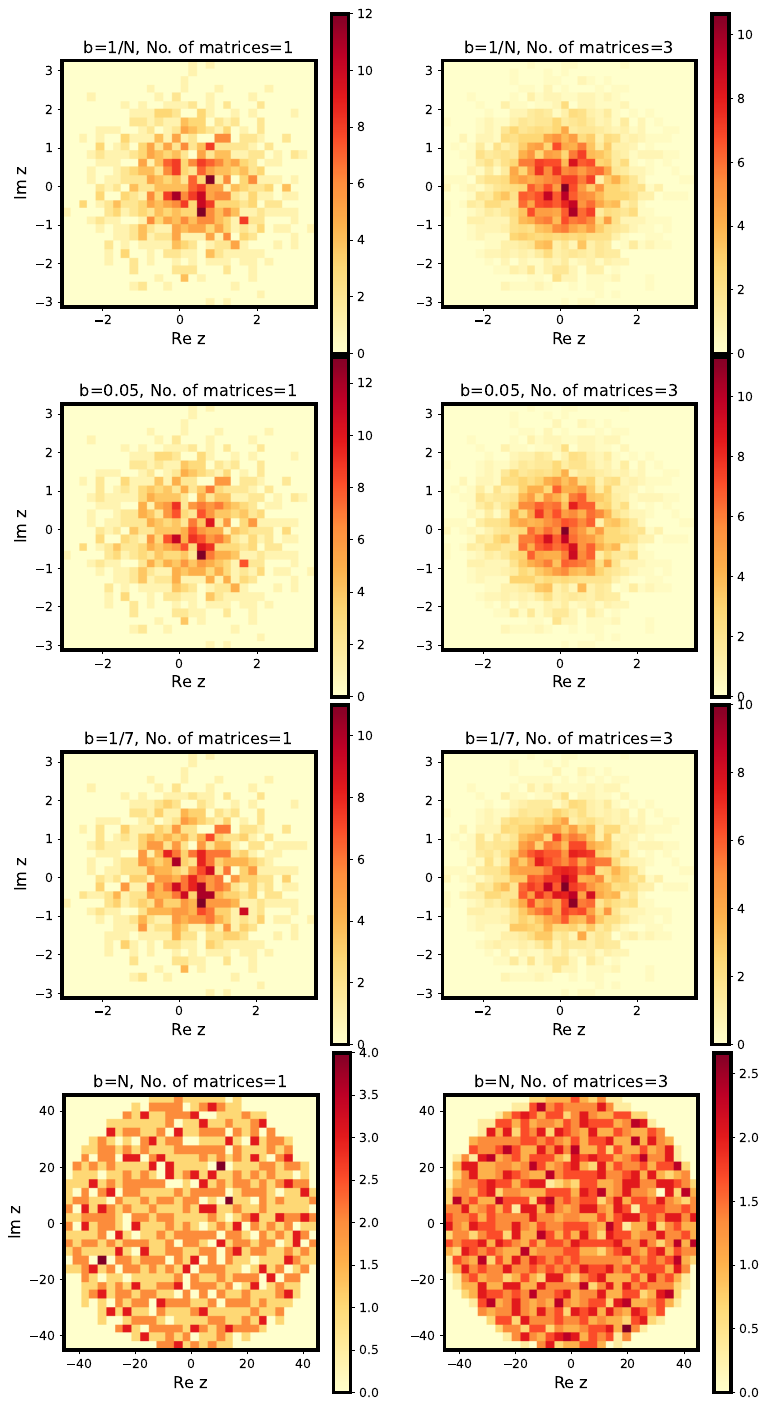}

\vspace{-0.8in}

\caption{{\bf Non-Ergodicity of the spectral density on the complex plane for BE :} The figure displays a comparison, for different $b$-values,  of  the  averaged spectral density for a single matrix (left panel) with that obtained by averaging  over three matrices (right panel).  As visuals indicate, the  two averaged densities appear more and more similar, thus implying an ergodic tendency, as $b$ increases. We note that, with increasing $b$, the off-diagonals in BE increasingly approach the diagonals, with $b \to \infty$ corresponding to a  Ginibre ensemble; the ergodicity of the spectral density in the latter case is already known.} 
\label{dbe}
\end{figure}

\begin{figure}[ht!]
\centering

\includegraphics[width=16cm,height=22cm]{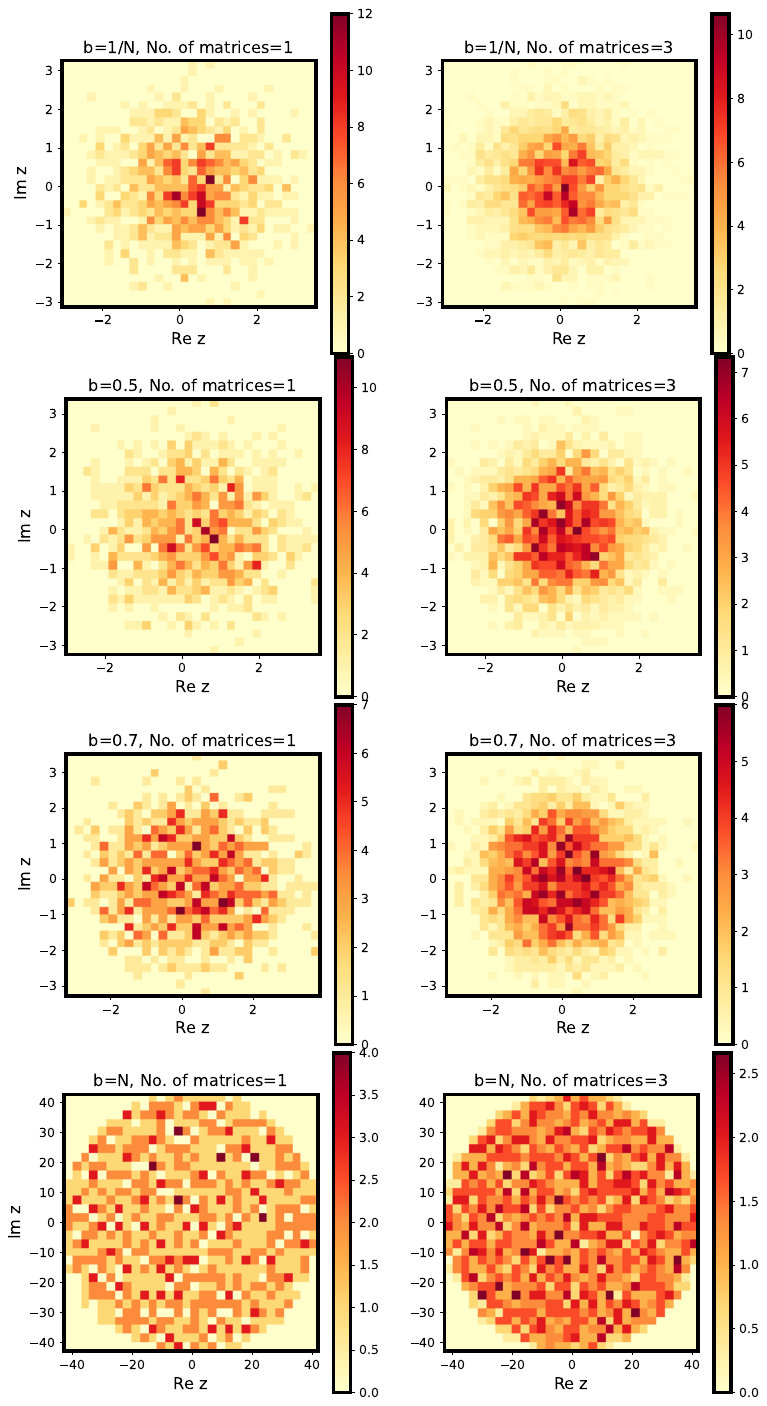}

\vspace{-0.8in}

\caption{{\bf Non-Ergodicity of the spectral Density  on the complex plane for PE:  }  while the other  details are same as in figure 1, the approach to ergodicity with increasing $b$ is visibly different in this case.}
 \label{dpe}
\end{figure}

\begin{figure}[ht!]
\centering

\includegraphics[width=16cm,height=22cm]{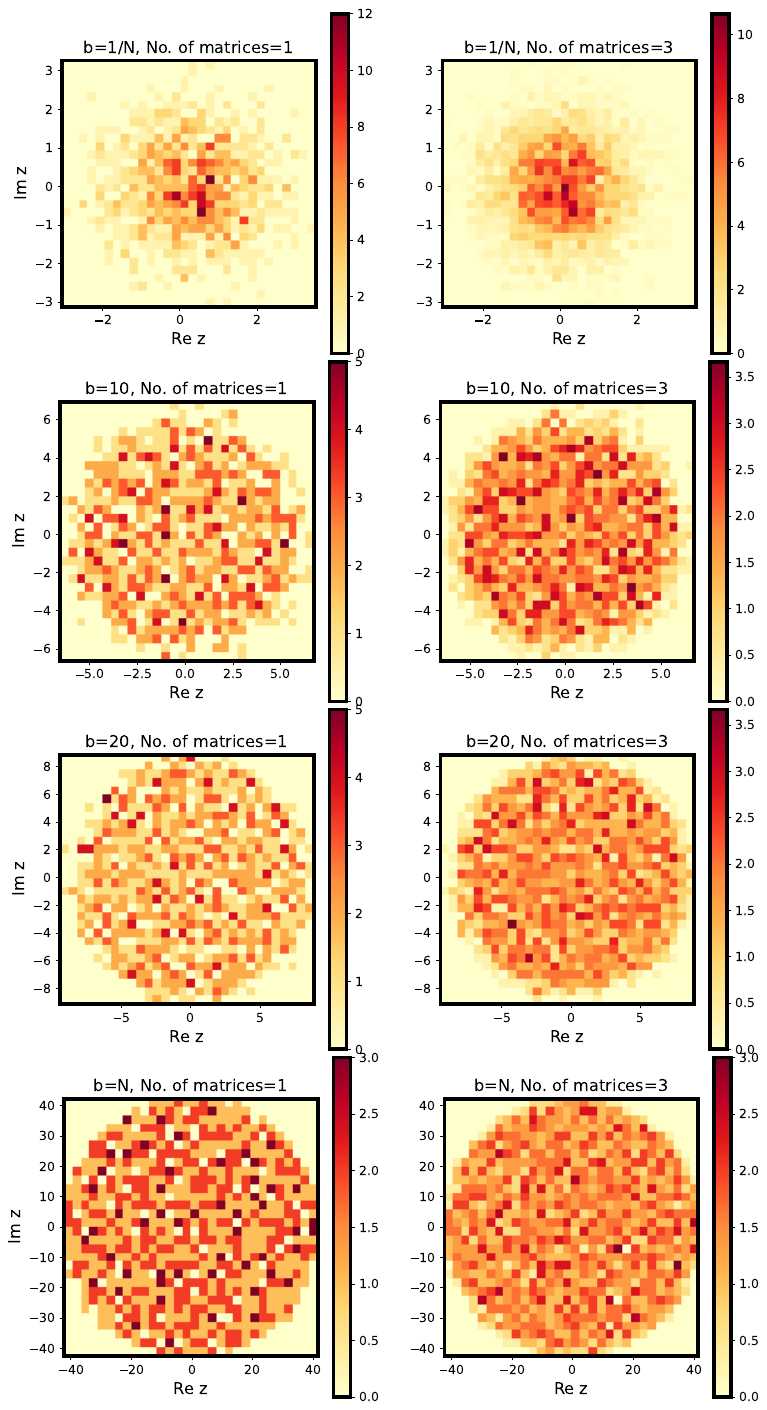}

\vspace{-0.8in}

\caption{{\bf Non-Ergodicity of the spectral Density on the complex plane for EE:  }  here again,  
 the approach to ergodicity with increasing $b$ is clearly different from that of BE and PE cases. The other  details are same as in figure 1}
\label{dee}
\end{figure}

\begin{figure}[ht!]
\centering

\vspace{-1.0in}

\includegraphics[width=20cm,height=25cm]{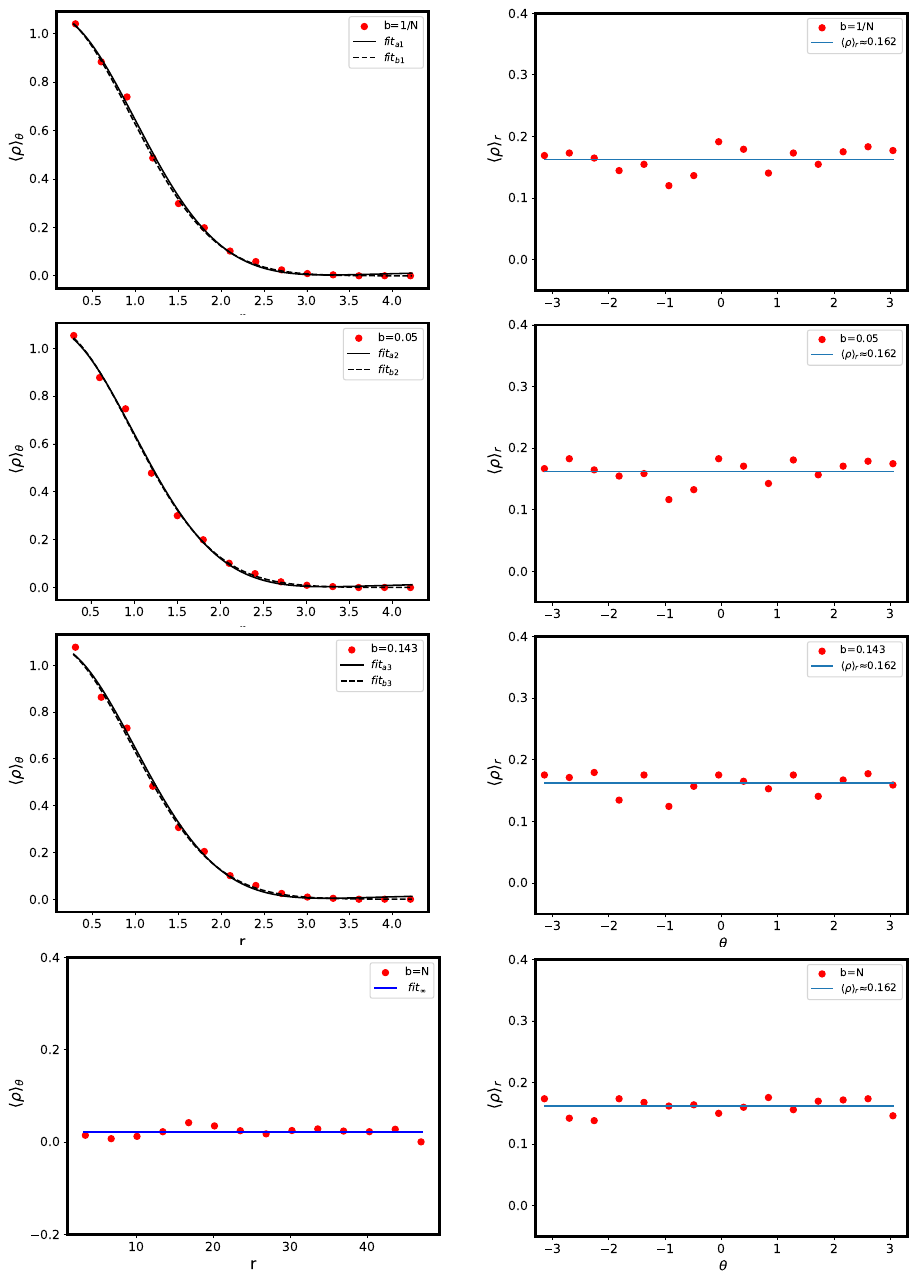}

\vspace{-1.3in}
\caption{{\bf Radial and angular dependence of the spectral density for BE:}  The figure displays the  radial  and angular dependence  (left and right columns and obtained by averaging over angle  and radial variables, respectively) of the ensemble averaged spectral density  $\langle \rho(r, \theta; Y) \rangle = N^{-1} \, R_1(r, \theta; Y)$ on the complex plane  for the BE for many $b$ values and fixed matrix size $N=1024$ and ensemble size ${\mathcal M}=10$.   The two fits  shown  for $b=1/N, 1/20, 1/7$ correspond to a comparison with eq.(\ref{rim2x}) amd eq.(\ref{rim2}).  The $(Y-Y_0)$ and the fitted functions for each $b$-value are given in table I. }

\label{fbe}
\end{figure}

\begin{figure}[ht!]
\centering

\vspace{-0.9in}

\includegraphics[width=20cm,height=25cm]{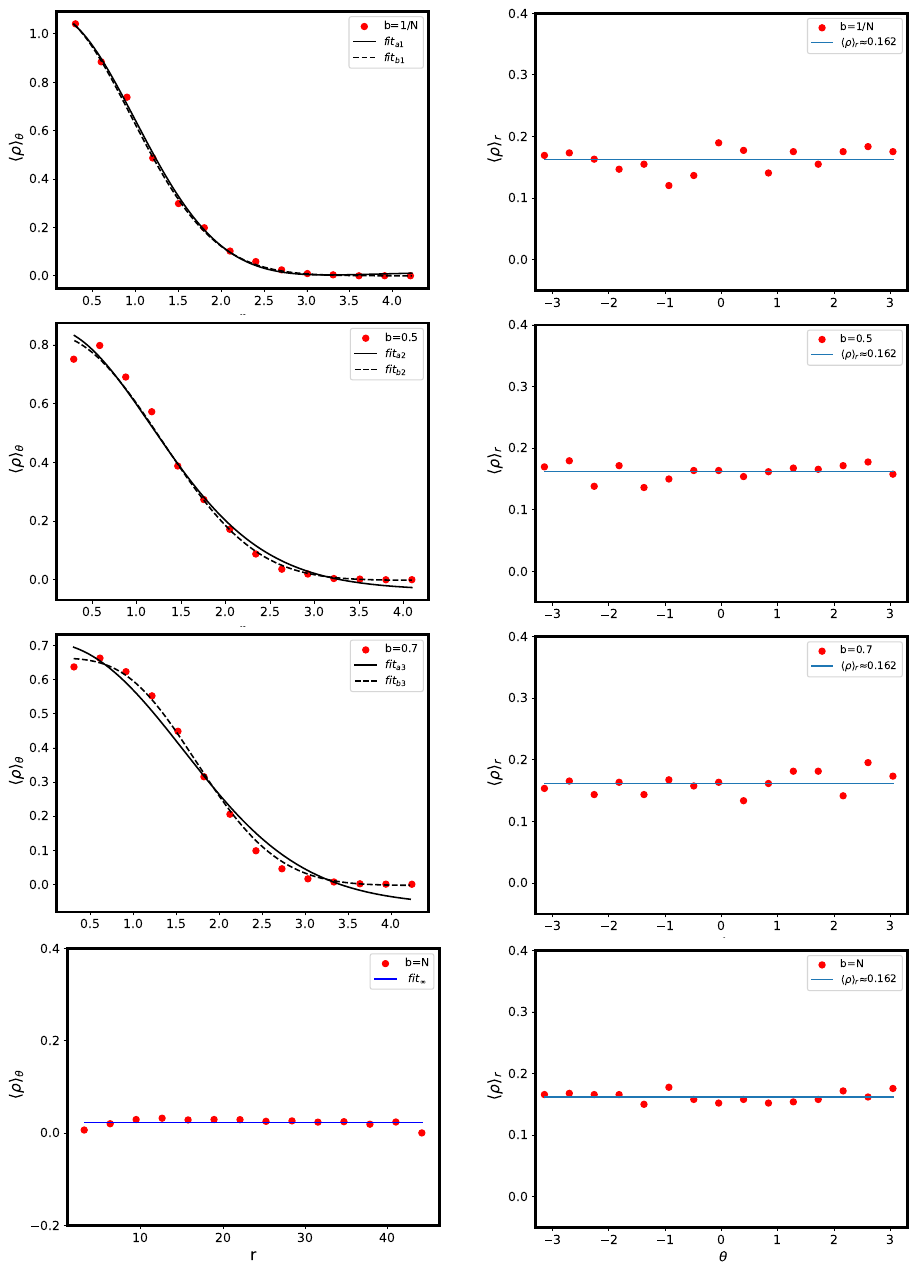}

\vspace{-1.2in}
\caption{{\bf Radial and angular dependence of the spectral density for PE:}  The figure displays  the  radial and angular dependence of the  ensemble averaged spectral density on the complex plane for many $b$ values and fixed $N=1024$ for PE.   The  $(Y-Y_0)$ values along with the details of fitted functions are given in table I.
Other details are same as in figure 4.}

\label{fpe}
\end{figure}

.   

\begin{figure}[ht!]
\centering

\vspace{-0.9in}

\includegraphics[width=20cm,height=25cm]{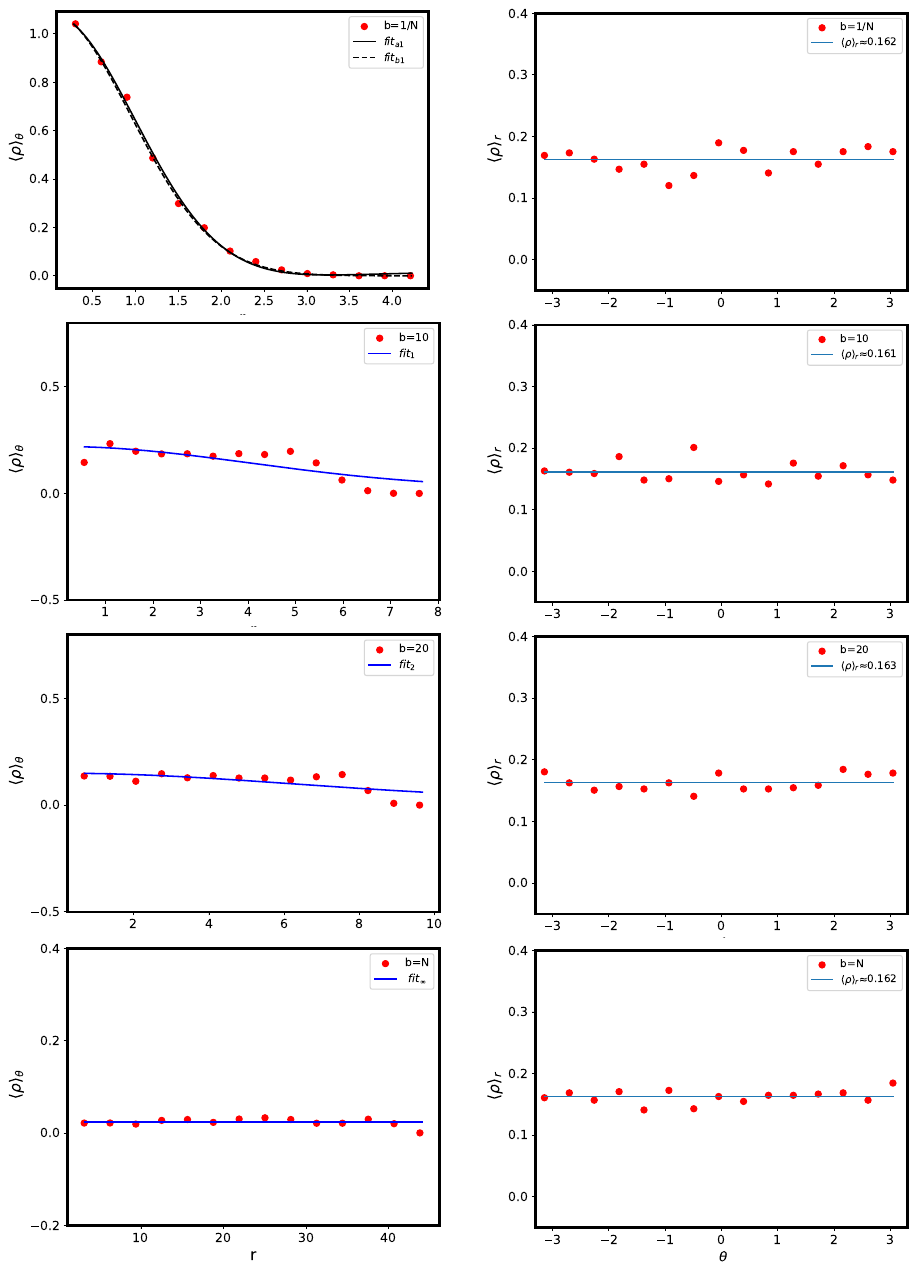}

\vspace{-1.2in}
\caption{{\bf Radial and angular dependence of the spectral density for EE:}  The figure displays  the  radial and angular dependence of the  ensemble averaged spectral density on the complex plane for many $b$ values and fixed $N=1024$ for the EE.   The  $(Y-Y_0)$ values along with  fitted functions in each case are given in table I; 
the  other details are same as in figure 4.  As in the case of BE and PE,  here again a smooth crossover from Poisson to Ginibre limit occurs as $b$ increases but the approach to a constant density  here is rapid for both radial as well as angle variables}

\label{fee}
\end{figure}

\newpage

\begin{table}
\caption{\bf Here we give the details of the fits used in figures 4-6 for BE, PE and EE respectively.
The mathematical expressions describing the three fits used are as follows:\\
$fit_{an}:\sum_{n=0}^{2}  q_{1n} \; U_{1n} \; e^{-2n\chi (Y-Y_0)}$,   for $n=1, 2, 3$, \\
$fit_{bn}:\frac{A}{\sqrt{r}}\; J_{1/2} (Br) \; e^{-C r^{2}}-2 \; \chi \; (Y-Y_0)\sum_{n=1}^2 \; n \; q_{1n} \; U_{1n}$,   for $n=1, 2, 3$, \\
$fit_n: (c_{10}+ d_{10} \, (\frac{r}{\sqrt{N}})^{N-1}e^{-r^{2}/2}) + (c_{11} \, e^{-C r^{2}}+ d_{11} (\frac{r}{\sqrt{N}})^{N-1}e^{-D r^{2}}) $,   for $n=1, 2, \infty$ \\
}

\begin{tabular}{|c|c|p{16cm}|}
\hline
\hline

{\bf b}           &         $\mathbf{Y-Y_0}$           &        { \bf Fits  for BE case (figure 4) }\\

\hline
       
1/N      &  0.00     &   \small{$fit_{a1}:q_{10}=132941.72,q_{11}=-266481.346,q_{12}=133540.709,\chi=0.001$}  \\&& \small{$fit_{b1}:A=1.952,B=0.493,C= 0.5, q_{11}=q_{12}=0 $}\\
\hline
1/20   & 7.88      &   \small{$fit_{a2}:q_{10}=137691.47,q_{11}=-280375.63,q_{12}=142729.85,\chi=0.001$}                    \\&& \small{$fit_{b2}:A=1.952,B=0.493,C= 0.5, q_{11}=-77.376, q_{12}=38.616, \chi=0.001$} \\

\hline
1/7           &  9.98     &  \small{$fit_{a3}:q_{10}=140349.813,q_{11}=-286984.046,q_{12}=146705.02 ,\chi=0.001    $} \\&& \small{$fit_{b3}:A=1.952,B=0.493,C= 0.5, q_{11}=-38.366, q_{12}=19.063, \chi=0.001 $} \\
\hline
N            & 26.34       &     \small{$fit_{\infty}:c_{10}= 0.021 ,d_{10} =2.589, c_{11}=d_{11}=0 $   }   \\
\hline
\hline

\hline
\hline
{\bf b}           &         $\mathbf{Y-Y_0}$           &        { \bf Fits  for PE case (figure 5) } \\
\hline
       
1/N               &  0.00 &  \small{$fit_{a1}:q_{10}=133214.964,q_{11}=-267028.282,q_{12}=133814.40,\chi=0.001$}  \\&& \small{$fit_{b1}:A=1.952,B=0.493,C= 0.5$ }  \\
\hline
0.5              & 12.46      &    \small{$fit_{a2}:q_{10}-90560.656,q_{11}=185602.35,q_{12}=-95096.20,\chi=0.001$}                    \\&& \small{$fit_{b2}:A=1.952,B=0.493,C= 0.5, q_{11}=-0.034, q_{12}=0.024, \chi=0.7 $ }\\
\hline
0.7           &  13.12    &  \small{$fit_{a3}:q_{10}=-227232.145,q_{11}=+466821.25,q_{12}=-239756.269,\chi=0.001$}                    \\&& \small{$fit_{b3}:A=1.952,B=0.493,C= 0.5, q_{11}=-0.055, q_{12}=0.039, \chi=0.7$ }\\
\hline
N            & 24.46      &     \small{$fit_{\infty}:c_{10}= 0.023,d_{10}=1.530, c_{11}=d_{11}=0 $}          \\
\hline
\hline
\hline
\hline 
{\bf b}           &         $\mathbf{Y-Y_0}$           &        { \bf Fits  for EE case (figure 6) }\\
\hline
       
1/N         &  0.00   &  \small{$fit_{a1}:q_{10}=133214.96,q_{11}=-267028.28,q_{12}=133814.40,\chi=0.001$}  \\&& \small{$fit_{b1}:A=1.952,B=0.493,C= 0.5$} \\
\hline
10              & 9.16E10      &    \small{$fit_{1}:c_{10}=0.024,d_{10}=1.0,c_{11}=0.197, d_{11}=0.15,  C=0.031,   
D= 0.046 $  }                  \\
\hline
20           &  9.16E10     &  \small{ $fit_{2}:c_{10}=0.024,d_{10}=1.0,c_{11}=0.127, d_{11}=0.249,  C=0.013, D= 0.019    $ } \\
\hline
N            & 9.16E10       &     \small{$fit_{\infty}: c_{10}= 0.024,d_{10}=1.442, c_{11}=d_{11}=0 $   }       \\
\hline
\hline

\end{tabular}
\end{table}



\begin{references}


\bibitem{berry} M. V. Berry, Czechoslovak J.Phys. 54, 1039, (2004).



\bibitem{r1}  D.  S.  Borgnia,  A.  J.  Kruchkov,  and  R.-J.  Slager,  ,  Phys.  Rev.Lett.124, 056802 (2020).
\bibitem{r2}  N.  Okuma,  K.  Kawabata,  K.  Shiozaki,  and  M.  Sato,  Phys.Rev. Lett.124, 086801 (2020).
\bibitem{r3}  S.  Yao  and  Z.  Wang, Phys. Rev. Lett.121,086803 (2018).
\bibitem{r4}  M. S. Rudner and L. S. Levitov,  Phys. Rev. Lett.102,065703 (2009).
\bibitem{r5}  Y. C. Hu and T. L. Hughes, Phys. Rev. B84, 153101 (2011).
\bibitem{r6}  K. Esaki,  M. Sato,  K. Hasebe, and M. Kohmoto, Phys. Rev. B84, 205128 (2011).
\bibitem{r7}  Z.  Gong,  Y.  Ashida,  K.  Kawabata,  K.  Takasan,  S.  Hi-gashikawa,  and  M.  Ueda,  Phys. Rev. X8, 031079 (2018).
\bibitem{r8}  H. Schomerus,  Opt. Lett.38, 1912 (2013).
\bibitem{r9}  Y.   Ashida,   Z.   Gong,   and   M.   Ueda,    Advances in Physics 69, 249 (2020).
\bibitem{r10}  N.Moiseyev, Non-Hermitian Quantum Mechanics (Cambridge University Press, 2011).
\bibitem{r11}  B.  Skinner,  J.  Ruhman,  and  A.  Nahum, Phys. Rev. X9, 031009 (2019).
\bibitem{r12}  A.  Zabalo,  M.  J.  Gullans,  J.  H.  Wilson,  R.  Vasseur,A. W. W. Ludwig, S. Gopalakrishnan, D. A. Huse, and J. H. Pixley,   Phys.  Rev.Lett. 128, 050602 (2022).
\bibitem{r13}  A.   C.   Potter   and   R.   Vasseur,  arXiv  e-prints, arXiv:2111.08018 (2021), arXiv:2111.08018 
\bibitem{r14}  J. Feinberg and A. Zee,  Phys. Rev. E59, 6433 (1999).
\bibitem{r15}  L. G. Molinari,  Journal of Physics A: Mathematical and Theoretical 42, 265204 (2009).
\bibitem{r16}  Y.  Huang  and  B.  I.  Shklovskii,  Phys. Rev. B 101, 014204 (2020).
\bibitem{r17}  K. Kawabata and S. Ryu,  Phys. Rev. Lett.126, 166801(2021).
\bibitem{r20}  G. L. Celardo, M. Angeli, and R. Kaiser, arXiv preprint arXiv:1702.04506  (2017).
\bibitem{r21}  F. Cottier, A. Cipris, R. Bachelard, and R. Kaiser,  Phys. Rev. Lett.123, 083401 (2019).
\bibitem{r22}  C.   E.   Maximo,    N.   A.   Moreira,    R.   Kaiser,   and R.  Bachelard, Phys. Rev. A100, 063845 (2019).
\bibitem{r23}  A.  Guo,  G.  J.  Salamo,  D.  Duchesne,  R.  Morandotti, M. Volatier-Ravat, V. Aimez, G. A. Siviloglou, and D. N.Christodoulides,   Phys.  Rev.  Lett.103,093902 (2009).

\bibitem{sc}
H.J.Sommers,   A. Crisanti, H. Sompolinsky and Y. Stein, Phys. Rev. Lett., 60, 1895 (1988).

\bibitem{ns}
D.R.Nelson and N.M.Shnerb,  58,  1383, (1998).


\bibitem{kr1}
K. Ranjan and L. F. Abbott, Phys. Rev. Lett., 97, 188104, (2006).

\bibitem{ahn}
A.  Amir, N. Hatano and D. R. Nelson, Phys. Rev. E 93, 042310, (2016).

\bibitem{afm}
Y. Ahmadian, F. Mumarola and K. D. Miller, Phys. Rev. E 91, 012820, (2015).


\bibitem{fte}
N. Hatano and D.R.Nelson, Phys. Rev. Lett. 77, 570 (1996);
K.B. Efetov, Phys Rev. B 56 9630 (1997);
I.Y. Glodsheild and B.A. Khoruzhenko, Phys. Rev. Lett., 80,
2897 (1998);
C.Mudry, B.D.Simons and A.Altland, Phys. Rev. Lett., 80,
4257 (1998).

\bibitem{cw}
J.T.Chalker and Z.J.Wang, Phys. Rev. Lett. 79, 1797, (1997).

\bibitem{cb}
J.T.Chalker and B.Mehlig, Phys. Rev. Lett. 81, 3367, (1998).

\bibitem{psnh}
P. Shukla, Phys. Rev. Lett. 87, 19, 194102, (2001).

\bibitem{ben1} C.M. Bender, N. Hassanpour, D.W.Hook, S.P.Klevansky, C. Sunderhauf and Z. Wen, Phys. Rev. A 95, (2017).

\bibitem{ben0} C.M. Bender and S. Boettcher, Phys. Rev. Lett. 80, 5243, (1998).

\bibitem{ben2}
C.M. Bender,  D.C. Brody and H.F.Jones, Phys. Rev. Lett. 89, (2002).

\bibitem{most}
A. Mostafazadeh, J. Math. Phys. 43, 205, (2002); 43, 2814, (2002); 43, 3944, (2002).

\bibitem{jk}
Y.N.Joglekar and W.A. karr,  Phys. Rev. E, 83, (2011)

\bibitem{gnt}
E-M Graefe, S. Mudute-Ndumbe and M. Taylor, J. Phys. A: Math. Theo. 48, 38FT02, (2015).




\bibitem{ir1}
I. Rotter, J. Phys. A: Math. Theo., 42, 153001, (2009).

\bibitem{ir2}
I. Rotter, arXiv:1011.0645, (2010).


\bibitem{gin}
J.  Ginibre,  J. Math. Phys. 6 440, (1965).


\bibitem{fh1}
F.Haake et al.,  Z.Phys. B, 88, 359, (1992).


\bibitem{nh}
Nils Lehmann and H-j Sommers, Phys. Rev. Lett. 67, 941, (1991).

\bibitem{fy1}
Y.V.Fyodorov and H.-J. Sommers, J. Math. Physics (N.Y.), 38, 1918, (1997).


\bibitem{fy2}
Y.V.Fyodorov,  B.A.Khoruzhenko and H.-J. Sommers, Phys. Rev. Lett.,
79, 557, (1997).

\bibitem{fy3}
Y.V.Fyodorov and B.A.Khoruzhenko, Ann. Inst. Henri Poincare (Physique
Theorique), 68, 449, (1998).


\bibitem{fz}
J. Feinberg and A.Zee,  Phys. Rev. E 59, 6433, (1999)

\bibitem{prosen}
L. Sa, P. Ribciro and T. Prosen, Phys. Rev. X, 10, 021019, (2020).

\bibitem{hkku}
R. Hamazaki, K. Kawabata, N. Kura and M. Ueda, Phys. Rev. Res., 2, 023286, (2020).

\bibitem{tk}
G.De Tomasi and I. Khaymovich, Phys. Rev. B, 106, 094204 (2022)

\bibitem{bp}
O. Bohigas and M. P. Pato, J. Phys. A: Math. Theor. 46, 115001,  (2013).

\bibitem{gnv}
A. M. Garcia-Garicia, S. M. Nishigaki and J.J.M. Verbaarschot,  Phys. Rev. E 66, 016132, (2002).


\bibitem{swhjy}
K. Suthar, Y-C Wang, Y-P Huang, H.H.Jens and J-H You,  Phys. Rev. B 106, 064208 (2022).



\bibitem{mcn}
A. M.  Mambuca,  C.  Cammarota and I. Neri,  Phys. Rev.  E 105, 014305 (2022). 


\bibitem{nm}
I. Neri and F.  Lucas Metz,  Phys.  Rev.  Research,  2, 033313 (2020).




\bibitem{haak} F.  Haake, Quantum Signatures of Chaos (Springer-Verlag,
Berlin, 1991).

\bibitem{meta}
M. L. Mehta,  Random Matrices (Academic, New York, 1991).



\bibitem{fkpt5}
J.B. French, V.K.B. Kota, A. Pandey and S. Tomsovic, Annals of Physics, 181, 198 (1988).

\bibitem{gmw}
T. Guhr, G A Muller-Groeling and H A Weidenmuller 1998 Phys. Rep. V 299 189

\bibitem{psrev}
P. Shukla,  Int. Jou. of Mod.Phys B (WSPC), 26, 12300008, (2012).



\bibitem{apbe}
A. Pandey, Chaos, Soliton and Fractals, 5, 1275, (1995).

\bibitem{apce}
A. Pandey,  Phase Transitions (Taylor and Francis), 77, 835 (2004).


\bibitem{brody}
T A Brody,  J Flores,  J B French,  P A Mello,  A Pandey and S S M Wong,  Rev.  Mod. Phys. 53 385 (1981).



\bibitem{psco}
P.Shukla,   Phys. Rev. E 71, 026266 (2005).

\bibitem{psalt}
P.Shukla,  Phys. Rev. E 62, 2098, (2000); 

\bibitem{psall}  
P. Shukla,  J. Phys.: Condens. Matter 17, 1653 (2005); J. Phys. A 41, 304023 (2008); Phys. Rev.  E 75, 051113, (2007); J. Phys. A: Math. Theor 50, 435003 (2017).
  
\bibitem{psall1}  
R. Dutta and P. Shukla, Phys. Rev. E, 76, 051124 (2007); 78, 031115 (2008); 
M. V. Berry and P.Shukla, J. Phys. A 42, 485102 (2009).
 S. Sadhukhan and P. Shukla, Phys. Rev. E 96, 012109 (2017);
P. Shukla, New J. Phys. 18, 021004 (2016).
  .

\bibitem{ptche}
 T. Mondal and P.Shukla, Phys Rev. E 102, 032131 (2020)
  

\bibitem{psacu}
P.Shukla,  J. Phys. A 41, 304023 (2008).



\bibitem{psflat}
P. Shukla, Phys. Rev. B, 98, 054206 (2018).


\bibitem{mnr}
F.L.Metz, I. Neri and T. Rogers,  J. Phys. A: Math. Theor. 52, 434003, (2019).


\bibitem{am} I. I. Arkhipov and F. Minganti, Phys. Rev. A, 107, 012202 (2023)

\bibitem{psgs2}
M. G. Ansari and P.  Shukla,  arXiv:2312.08203 [cond-mat.dis-nn].


\bibitem{circ}
A.Pandey and P.Shukla, J. Phys. A, (1991).


\bibitem{temme}
N.M. Temme and E.J.M. Veling,   Indagationes Mathematicae  (Elsevier) 33, 1221, (2022).



\bibitem{psps}
Although at present we are not aware of any real physical non-Hermitian operator with phase singularity at  $r=0$ on spectral plane ,  but such a situation can be envisaged to arise,  for example,  for a system with vanishing spectral density $\rho(r, \theta)$ at $r=0$.  




\end{references}
\end{document}